\documentclass[sn-mathphys,Numbered]{sn-jnl}

\usepackage{graphicx}%
\usepackage{multirow}%
\usepackage{amsmath,amssymb,amsfonts}%
\usepackage{amsthm}%
\usepackage{mathrsfs}%
\usepackage[title]{appendix}%
\usepackage{xcolor}%
\usepackage{textcomp}%
\usepackage{manyfoot}%
\usepackage{booktabs}%
\usepackage{algorithm}%
\usepackage{algorithmicx}%
\usepackage{algpseudocode}%
\usepackage{listings}%

\pdfoutput=1
\theoremstyle{thmstyleone}%
\theoremstyle{thmstyletwo}%

\theoremstyle{thmstylethree}%

\begin{document}

\title[Article Title]{SdCT-GAN: Reconstructing CT from Biplanar X-Rays with Self-driven Generative Adversarial Networks}


\author[1]{\fnm{Shuangqin} \sur{Cheng}}\email{sqcheng@stu2021.jnu.edu.cn}

\author*[1]{\fnm{Qingliang} \sur{Chen}}\email{tpchen@jnu.edu.cn}
\equalcont{These authors contributed equally to this work.}

\author[1]{\fnm{Qiyi} \sur{Zhang}}\email{qiyizhang@stu2021.jnu.edu.cn}

\author[1]{\fnm{Ming} \sur{Li}}\email{liming9653@gmail.com}

\author[2]{\fnm{Yamuhanmode} \sur{Alike}}\email{yamuhanmode@mail.sysu.edu.cn}

\author[3]{\fnm{Kaile} \sur{Su}}\email{k.su@griffith.edu.au}

\author[4]{\fnm{Pengcheng} \sur{Wen}}\email{lnt1415926@stu2019.jnu.edu.cn}

\affil*[1]{\orgdiv{School of Information Science and Technology}, \orgname{Jinan University}, \orgaddress{\street{Shipai Street, Tianhe District}, \city{Guangzhou}, \postcode{510632}, \state{Guangdong}, \country{China}}}

\affil[2]{\orgdiv{orthopedics}, \orgname{Sun Yat-sen Memorial Hospital of Sun Yat-sen University}, \orgaddress{\street{No.103 Yanjiang West Road}, \city{Guangzhou}, \postcode{510632}, \state{Guangdong}, \country{China}}}

\affil[3]{\orgdiv{Integrated and Intelligent Systems}, \orgname{Griffith University}, \orgaddress{\street{No.103 Yanjiang West Road}, \city{Nathan}, \postcode{4111}, \state{}, \country{Australia}}}

\affil[4]{\orgdiv{International School}, \orgname{Jinan University}, \orgaddress{\street{Shipai Street, Tianhe District}, \city{Guangzhou}, \postcode{510632}, \state{Guangdong}, \country{China}}}


\abstract{Computed Tomography (CT) is a medical imaging modality that can generate more informative 3D images than 2D X-rays. However, this advantage comes at the expense of more radiation exposure, higher costs, and longer acquisition time. Hence, the reconstruction of 3D CT images using a limited number of 2D X-rays has gained significant importance as an economical alternative. Nevertheless, existing methods primarily prioritize minimizing pixel/voxel-level intensity discrepancies, often neglecting the preservation of textural details in the synthesized images. This oversight directly impacts the quality of the reconstructed images and thus affects the clinical diagnosis. To address the deficits, this paper presents a new self-driven generative adversarial network model (SdCT-GAN), which is motivated to pay more attention to image details by introducing a novel auto-encoder structure in the discriminator. In addition, a Sobel Gradient Guider (SGG) idea is applied throughout the model, where the edge information from the 2D X-ray image at the input can be integrated. Moreover, LPIPS (Learned Perceptual Image Patch Similarity) evaluation metric is adopted that can quantitatively evaluate the fine contours and textures of reconstructed images better than the existing ones. Finally, the qualitative and quantitative results of the empirical studies justify the power of the proposed model compared to mainstream state-of-the-art baselines.}




\maketitle

\begin{figure}[!ht]
  \centering
  \includegraphics[width=\columnwidth]{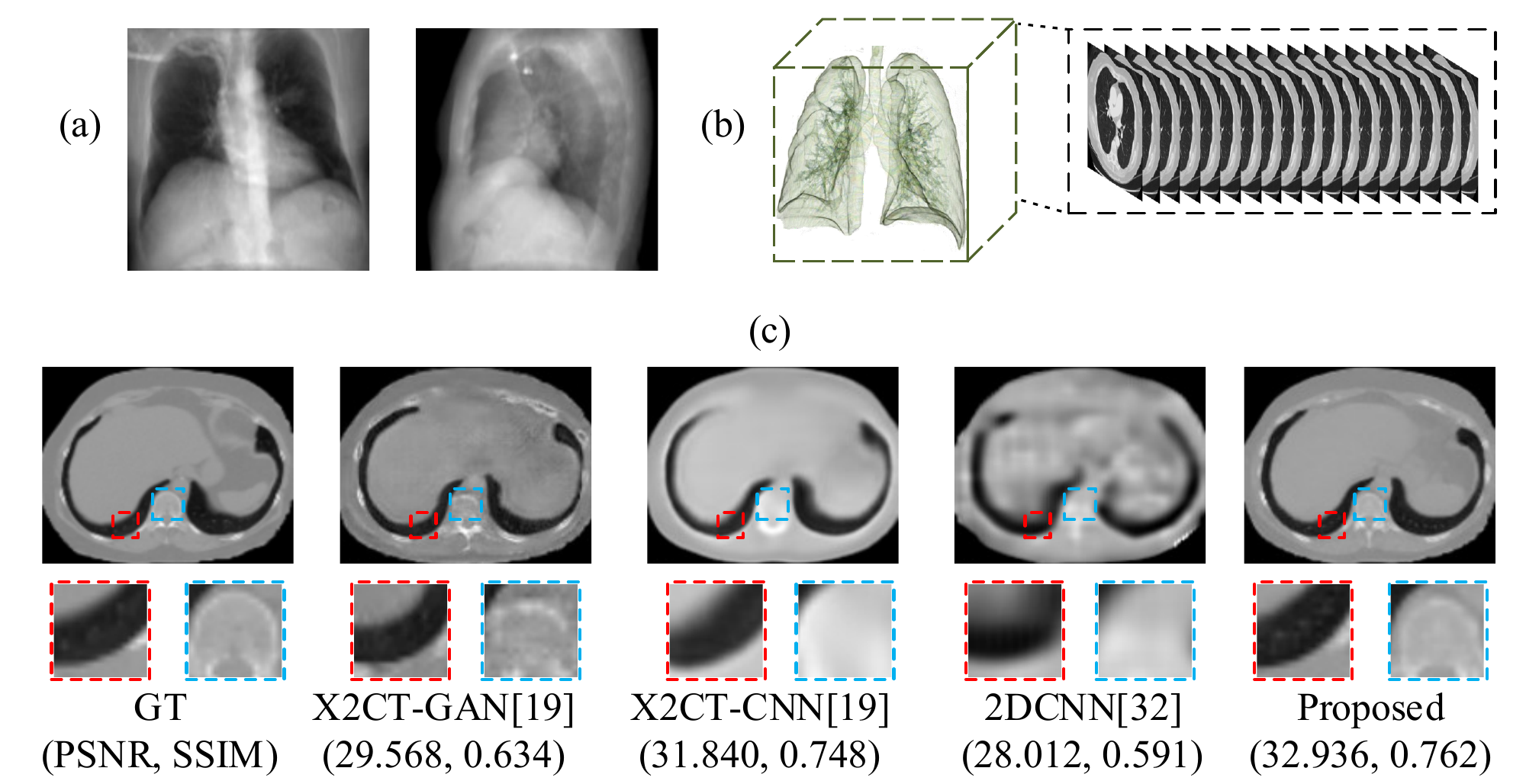}
  \caption{(a) Two orthogonal X-ray sheets as input (LIDC-IDRI-0561), (b) Our proposed method successfully recovers the 3D density field of the patient, including certain texture details and geometric structures in the CT tomography. (c) PSNR and SSIM metrics are employed for comparing our results with state-of-the-art (SOTA) methods. The 2DCNN method, a deep convolutional neural network, exhibits  notable blurring of both geometric boundaries and fine details in the reconstructed images. By contrast, X2CT-GAN, a generative adversarial network method with multiple conditional constraints, has difficulty in accurately reconstructing intricate details. Although X2CT-CNN, a generative adversarial network method constrained by MSE loss only, demonstrates superior performance in terms of PSNR and SSIM metrics compared to X2CT-GAN, it struggles to capture image details, as evidenced by the unsatisfactory results highlighted in the red and blue boxes.}\label{fig:first_imag}
\end{figure}

\begin{figure*}[!t]
  \centering
  \includegraphics[width=0.9\textwidth]{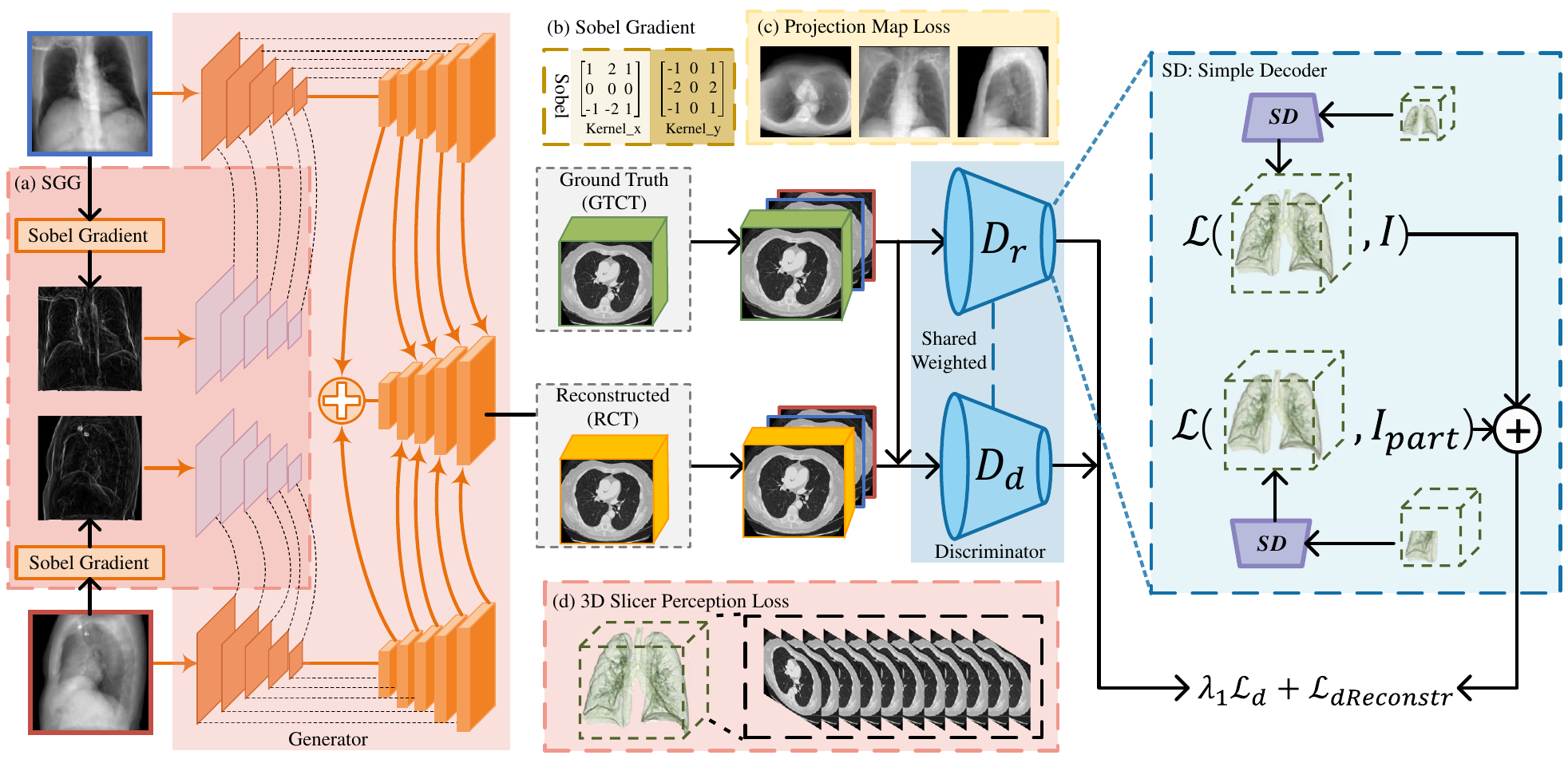}
  \caption{The SdCT-GAN framework comprises a generator G and two discriminators, $D_{r}$ and $D_{d}$. The generator consists of two parallel encoder-decoder networks based on the U-Net structure, a fusion network, and the proposed SGG module to enhance edge information during encoding. The discriminator $D_{d}$ is trained to differentiate between GTCT and RCT triplets, while the discriminator$D_{r}$ employs a simple decoder structure to establish reconstruction loss constraints $L_{dReconstr}$ with GTCT inputs, enabling it to extract global and local features more effectively. For a detailed diagram of the discriminator D framework, refer to {Figure~3}. The generator is trained to synthesize RCT that closely resemble the GTCT, while the discriminator is trained to discern between the authenticity of the RCT and GTCT. The final result is a highly restored reconstructed image through their adversarial interplay}\label{fig: SdCT-GAN-frame}
\end{figure*}

\section{Introduction}\label{sec1}

tomography (CT) has become a popular imaging technology for various diseases due to its excellent visualization of internal structures\cite{taguchi2013vision}. Conventional CT imaging acquires X-ray attenuation measurements from multiple angles to generate a sinogram\cite{hounsfield1995computerized}, which is subsequently reconstructed using techniques like Filtered Back Projection (FBP) or Algebraic Reconstruction Techniques (ART)\cite{gordon1970algebraic}. However, FBP is fast but susceptible to artifacts\cite{herman2009fundamentals}, while ART is iterative but computationally demanding\cite{gilbert1972iterative, beister2012iterative}. 
Besides, traditional CT reconstruction methods require a 360-degree scan of X-ray projections, resulting in significant radiation exposure to the human body\cite{padole2015ct}. To address these issues, researchers often employ reduced scanning intensity, 1/6 sparse angular sampling, or limited-angle scanning within the range of [0, 120°] to obtain CT images\cite{sidky2008image}. However, such approaches lead to severe streak artifacts and noises in low-dose CT images reconstructed from undersampled projections\cite{willemink2019evolution}. In recent years, various techniques have been proposed to obtain clearer CT images, primarily involving post-processing with filtering methods in the image domain\cite{geraldo2016low, yuan2018adaptive}, preprocessing with interpolation methods in the sinogram \cite{dong2019sinogram, anirudh2018lose}, and iterative reconstruction combined with deep learning\cite{pan2022multi,mizusawa2021computed,zang2021intratomo}. Despite the use of sparse or limited-angle CT, the radiation dose can still be over 50 times that of a radiograph, besides CT being more expensive and time-consuming\cite{coffey2010patient,mettler2012essentials}. 

 In recent years, deep learning has shown great potential in image reconstruction and recognition by leveraging large datasets and powerful neural networks\cite{ying2019x2ct, ratul2021ccx, jiang2022mfct, sun2022ultra, shen2022geometry}. 
Deep learning-based image reconstruction methods can overcome the limitations of traditional approaches by learning a mapping between the input 2D radiograph and 3D CT tomographic target images\cite{jin2017deep}. Despite the impressive performance of deep learning-based methods, there are still challenges ahead that need to be addressed in this area. Firstly, most existing methods\cite{ying2019x2ct,ratul2021ccx,shen2022geometry} only focus on minimizing pixel/voxel-wise intensity differences but ignore the textural details, which potentially affects the quality of synthesized images. Secondly, current CT reconstruction methods often rely on single quantitative assessments of image quality, such as PSNR (Peak Signal-to-Noise Ratio)  and SSIM (Structural SIMilarity)\cite{shen2022nerp,yu2019ea}. However, these metrics do not fully capture the perceptual quality of the reconstructed images, particularly in terms of internal details, depicted in {Figure~\ref{fig:first_imag}(c)}. Thus, there is a need to develop more comprehensive evaluation metrics that are aligned with human visual perception. Finally, since 2D X-ray data lacks depth information, most current methods rely on multiple views of the input X-ray data to recover 3D information. However, in real-world applications, only orthogonal X-rays, or even just a front view, are often available. Therefore, it is of practical significance to develop more efficient and robust reconstruction methods that can work with limited input views. 

In summary, the existing approaches are still far from satisfactory. In this paper, we address the challenges of optimizing deep learning reconstruction algorithms for CT reconstruction tasks using only 1-2 radiographs instead of sinograms to achieve high-quality CT imaging effects with lower X-ray radiation doses. Our task involves 2D-3D reconstruction, which is a more challenging task than 3D-3D reconstruction since it needs to recover three-dimensional information data from partial two-dimensional one\cite{chenes2021revisiting}. And furthermore, our network model can be generalized and extended to 3D-3D reconstruction tasks, and our approach can improve CT imaging while minimizing radiation exposure.

To be more specific, this paper proposes a new 3D self-supervised generative adversarial network model base on X2CT-GAN\cite{ying2019x2ct} that can reconstruct 3D CT from two orthogonal X-rays, with significant improvements on the enhancement of 3D image boundaries, depth texture information and contours. And the main contributions of the paper can be summarized as follows:

\begin{itemize}
    \item To the best of our knowledge, this is the first work to employ a novel automatic regularization discriminator framework for the task of 3D tomographic image synthesis, emphasizing both local details and global information of the images.
    \item 3D reconstruction image quality evaluation metric LPIPS (Learned Perceptual Image Patch Similarity) is adopted in comparative analyses, which better reflects the reconstruction quality of the detailed information inside the CT image and better matches the human visual perception.
    \item The final SdCT-GAN model shown in {Figure~\ref{fig: SdCT-GAN-frame}} has been evaluated and achieved the best experimental results in both qualitative and quantitative analysis in the public benchmark compared to other state-of-the-art baselines. Furthermore, unlike most previous approaches, our network does not require any image annotations or a priori knowledge for training. 
    \item Our code is available on GitHub.
\end{itemize}

\section{Related Work}\label{sec2}

\subsection{Tomographic Reconstruction in Medical Images}
Tomographic reconstruction in medical images can be divided into unimodal and multimodal depending on the input image type. The unimodal tomographic reconstruction task applies low-dose CT (LDCT) to synthesize artifact-free\cite{bera2021noise} or uses dynamic CT angiography to construct standard CT angiography images\cite{wu2022vessel}. And the multimodal tomographic reconstruction task can be further subdivided into 3D-3D and 2D-3D, respectively. In 3D-3D, MRI (Magnetic Resonance Imaging) is fed to generate CT images\cite{liu2021ct,oulbacha2020mri}, and some approach like NeRP\cite{shen2022nerp} exploits the internal information in an image prior and the physics of the sparsely sampled measurements to produce a representation of the unknown subject. 

Generally, 2D-3D reconstruction strives to recover 3D information from partially 2D ones, which is a rather challenging endeavour\cite{chenes2021revisiting}. Models based on the powerful deep convolutional neural networks (DCNNs) have thus been proposed to meet the challenge. Henzler et al\cite{henzler2018single} are the first to propose using deep convolutional neural networks to extend the 2D X-Rays image into a 3D volume. X23D\cite{jecklin2022x23d} introduces a new shape estimation method that extracts patient-specific 3D shape representations of the lumbar spine by using multi-view sparse X-ray data and corresponding image calibration parameters as input. It is worth emphasizing that the two aforementioned methods focus on reconstructing the 3D shape without considering the internal structure, whereas our method specifically addresses the internal structure of 3D cross-sections and is capable of rendering consistent 3D shapes as well.
ReconNet\cite{shen2019patient} shares a similar deep network structure with 2DCNN\cite{henzler2018single}, consisting of an encoder model, dimension transformation model, and decoder model. And the key difference between them lies in its usage. PatRecon utilizes a dataset generated from CT scans of a specific patient to train a patient-specific deep learning model, which can then be employed for subsequent CT imaging of the same patient. X-ray2CTNet\cite{sun2022ultra} introduces two-view feature adaptive fusion modules to construct a depth network model for cross-dimensional feature mapping from a 2D X-ray to a 3D volume.  However, their training sets are constructed by simulating different particle regression models for a certain type of propellant and using a CT simulation platform. Therefore it is difficult to predict the effect for application to real CT tomographic reconstructions. Shen et al\cite{shen2022geometry} establish a geometry-informed deep learning framework for ultra-sparse 3D tomographic image reconstruction. But labeling geometric a priori information for X-ray is a time-consuming and laborious task. By contrast, our approach does not require any image annotations or prior knowledge for training, making it universally applicable. Specifically, we place greater emphasis on the reconstruction of fine details within tomographic images, with comprehensive qualitative and quantitative analyses.

\begin{figure}[!t]
  \centering
  \includegraphics[width=\columnwidth]{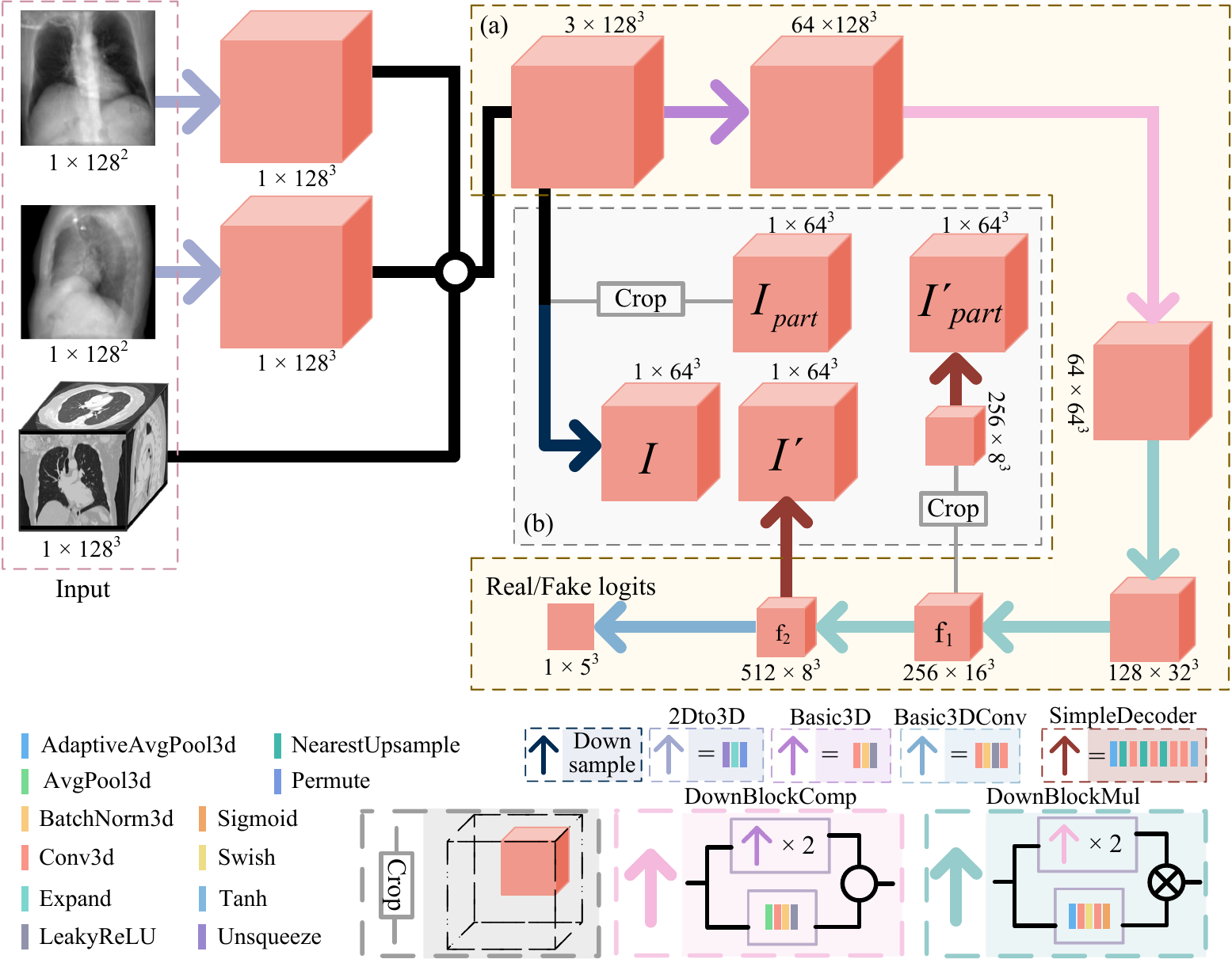}
  \caption{An overview of the proposed Discriminator with AutoEncoder (DAE). Only the real image can be simultaneously inputted into modules (a) and (b) in the discrimination process. When the real image enters module (a), it produces intermediate feature maps $f_1$ and $f_2$. These feature maps are then fed into the module (b), with $f_1$ being subjected to random cropping and Simple Decoder to obtain ${I^{'}}_{part}$, and $f_2$ directly transformed into ${I^{'}}$ through Simple Decoder. In module (b), the real image undergoes cropping and downsampling to generate $I_{part}$ and $I$, respectively. It is crucial to ensure that the cropping method and random position used for ${I^{'}}_{part}$ and $I_{part}$ are consistent with those used for $f_1$. By training the model, the differences between ${I^{'}}_{part}$ and $I_{part}$, as well as between ${I^{'}}$ and $I$, are compared to drive module (b) to extract more significant and fine-grained features. Both synthetic and real images can be inputted into module (b) to determine the authenticity of the images. }\label{fig:discrimintor}
\end{figure}

\subsection{Reconstruction of Textural Details}
The detailed structural information of 3D CT images is quite essential for generating high-quality CT. 
Generative Adversarial Networks (GAN), currently the most influential model for image fineness reconstruction, adopts the idea of a zero-sum game in game theory, in which the generative network G (Generator) and the discriminator network D (Discriminator) play continuously, and then G can thus be forced to learn the data distribution of the input. X2CT-GAN \cite{ying2019x2ct} is the latest prestigious model that is based on a new GAN network structure to generate CT from only two X-ray views. More recently, Ratul et al\cite{ratul2021ccx} try to encourage the model to focus on local details by prior semantic information. Therefore it needs additionally to train a semantic segmentation network model to get their conditional inputs, while our proposed approach is end-to-end. 

And during the interplay, preventing pattern collapse in the generator is crucial for ensuring stable GAN training. It has been emphasized in \cite{gulrajani2017improved, liu2020towards} that the discriminator (D) is susceptible to overfitting, leading to the lack of informative gradients for training the generator (G). Consequently, several contemporary approaches focus on stabilizing GAN training by addressing this issue from an optimal discriminator standpoint. Fan et al\cite{fan2022tr} propose a multi-scale and positional pixel discriminator that incorporates partition mixing and random cropping techniques. This approach aims to enhance the network's learning capacity by attaching importance to discriminated regions and local features. 
Yu\cite{yu2019ea} and Jiang et al\cite{jiang2021synthesis} input the Sobel feature gradient map to the discriminator with the original image or the synthetic image concatenation, thus preserving the edges and alleviating the blurring problem caused by MSE (mean square errors).
Unlike them, our proposed approach do not need additional preprocessing (partitioning, cropping, edge extraction) on the synthetic or real images fed to the discriminator, and is independent of the data and the generator, focusing solely on the self-coding of the discriminator during training.

\begin{figure}[!t]
  \centering
  \includegraphics[scale=0.8]{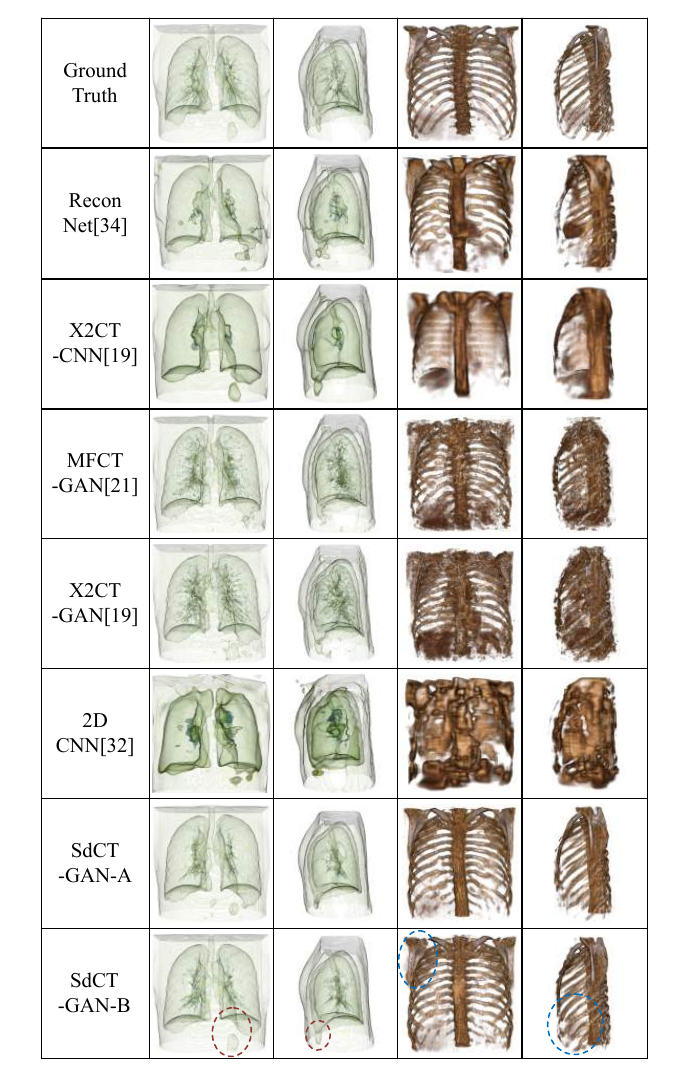}
  \caption{Reconstructed CT scans with different approaches. The first and second columns show the 3D effects of the generated frontal and lateral views of the CT lung texture, respectively. The third and fourth columns exhibit the 3D effects of the generated CT frontal and lateral views of the chest bones, respectively. SdCT-GAN-A is the proposed SdCT-GAN with the reconstruction loss ${L}_{dReconstr}$ in the discriminator containing only the voxel reconstruction loss function ${L}_{dVoxel}$.  SdCT-GAN-B is that the reconstruction loss in the discriminator contains a voxel reconstruction loss function ${L}_{dVoxel}$ and a perceptual reconstruction loss function ${L}_{d3DPecpt}$.}\label{fig:reconstructed_CT}
\end{figure}

\begin{figure}[!t]
  \centering
  \includegraphics[scale=0.48]{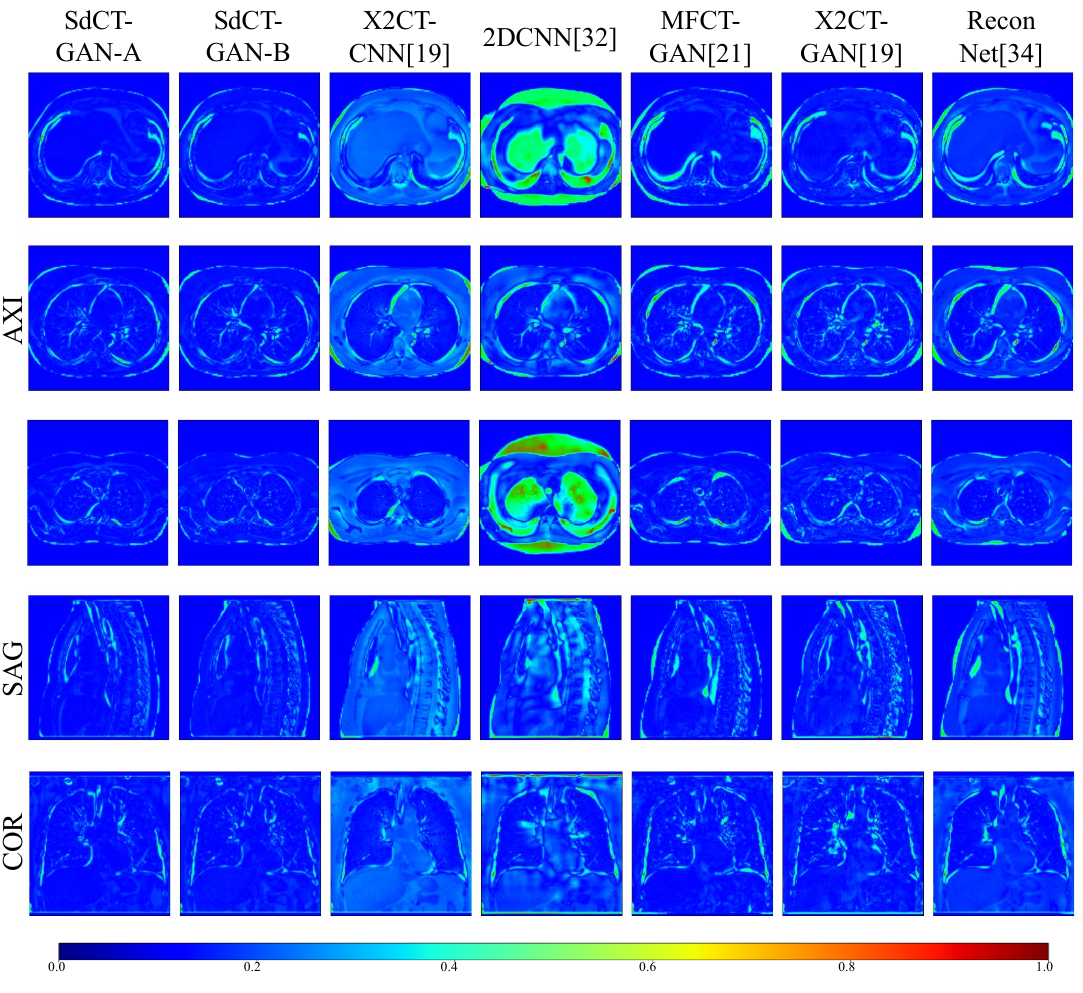}
  \caption{Comparison of error maps of CT images along sagittal (SAG), coronal (COR), and axial (AXI) planes. The pixel values of the original images range from 0 to 255. The error maps are constructed using $|y_{i} - \hat{y}_{i}|$, where the pixel values represent the differences between the ground truth and generated images. To better illustrate the results, the error map values are scaled to the range of [0, 1] by dividing them by 255. So, the pixel's color being closer to blue (0 in the color plot) is interpreted as the error being smaller. It can be seen from this visualization that the proposed SdCT-GAN outperforms the other baselines.}\label{fig:error_map}
\end{figure}

\begin{figure*}[!t]
\centering
\includegraphics[width=\columnwidth]{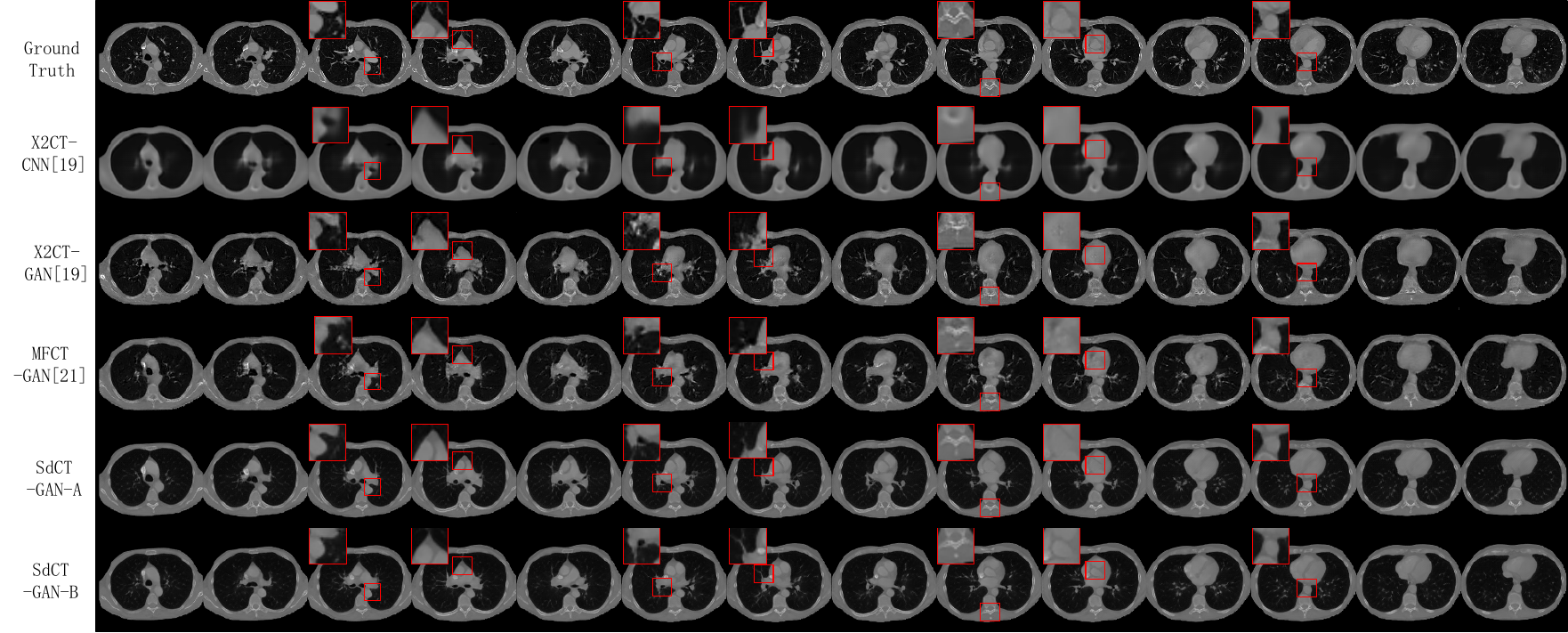}
\caption{The synthesis of 3D CT cross-sectional results with the proposed SdCT-GAN model by taking two orthogonal real X-rays (LIDC-IDRI-0007) as input after CycleGAN\cite{zhu2017unpaired, ying2019x2ct} transformation. 
}\label{fig:detail_map}
\end{figure*}

\section{Network Architecture}\label{sec3}

Now we present the proposed self-supervised learning generative adversarial network model
(SdCT-GAN), whose overall idea is that edge information is delicately integrated into the network architecture of the generator of SdCT-GAN and the 3D information of the reconstructed CT images is complemented by a novel auto-encoder discriminator.
\subsection{Generator}
The generators are designed for frontal and lateral X-rays, with two parallel encoder-decoder networks and a fusion network in the middle, fusing 3D fundamental blocks to incorporate discriminative information. 

\subsubsection{SGG (Sobel Gradient Guider) Module }
Edge information in medical images can well reflect the texture structure of the image content and describe the boundaries of different objects in the image\cite{GonzalezW08}. It has been shown in \cite{lu2010selective, wei2019synthesizing} that regularization of gradient types can preserve fine-grained feature structure. And Yu et al\cite{yu2019ea} propose gEa-GAN to use the difference between the edge images obtained from the Sobel filter as the objective function for gradient guidance. \cite{stimpel2019projection} has used the gradient map to weight the losses so that losses arising from edges are reinforced and losses arising from homogeneous regions are weakened. And GGGAN\cite{jiang2021synthesis} and dEa-GAN\cite{yu2019ea} are inputting Sobel feature maps into the discriminator. All of these methods are the process of conditionally constraining the synthetic image with the real image at the output side through objective function, adversarial learning, and then continuously fitting through backpropagation of the neural network. By contrast, our proposed SGG module is firstly to run the Sobel operator gradient on the original frontal and lateral X-rays in the generator input side to obtain the related gradient feature maps, which can be considered as image enhancements to the initial X-ray semantic types. The gradient feature map is then fed into the multi-layer perceptron in the SGG module. Finally, the intermediate features of the SGG module are merged with the encoder's middle features in the main line. As was pointed out in\cite{kim2016deeply}, if directly integrating the gradient feature map with the original picture, the operator gradient features will get lost as the number of layers in the network grows. However, this problem can be avoided with the SGG module.

\begin{table}[!ht]
\centering
\caption{Quantitative results (Mean ± Standard Deviation).In SdCT-GAN, the single view configuration refers to the generator consisting of a single encoder-decoder network, where the input is solely the X-ray image of the positive view.}\label{tab:quantitative_res}
\begin{tabular*}{\textwidth}{@{\extracolsep\fill}lcccccc}
\toprule
 & \textbf{Method} & \textbf{RSNR} & \textbf{SSIM} & \textbf{LPIPS} & \textbf{RNMSE(Hu)} \\
 \midrule
\multirow{7}{*}{\textbf{\begin{tabular}[c]{@{}c@{}}Single\\ -view\end{tabular}}} & \textit{2DCNN}\cite{henzler2018single} & 25.579±3.161 & 0.501±0.070 & 45.340±6.264 & 0.502±0.094 \\
 & \textit{ReconNet}\cite{shen2019patient} & 26.654±3.364 & 0.564±0.076 & 41.234±6.695 & 0.457±0.094 \\
 & \textit{X2CT-CNN}\cite{ying2019x2ct} & 27.476±3.446 & 0.586±0.075 & 43.001±5.881 & 0.423±0.097 \\ 
 & \textit{X2CT-GAN}\cite{ying2019x2ct} & 26.575±3.282 & 0.524±0.075 & 37.935±5.998 & 0.456±0.096 \\
 & \textit{MFCT-GAN}\cite{jiang2022mfct} & 26.705±3.420 & 0.534±0.079 & 36.402±6.185 & 0.456±0.101 \\
 & \textit{SdCT-GAN-A} & 27.251±3.320 & 0.554±0.070 & 36.889±6.074 & 0.435±0.093 \\
 & \textit{SdCT-GAN-B} & \textbf{27.379±3.391} & \textbf{0.566±0.076} & \textbf{35.746±6.199} & \textbf{0.437±0.100} \\ \hline
\multirow{7}{*}{\textbf{\begin{tabular}[c]{@{}c@{}}Two\\ -view\end{tabular}}} & \textit{2DCNN}\cite{henzler2018single} & 28.012±2.512 & 0.591±0.052 & 42.692±5.935 & 0.387±0.068 \\
 & \textit{ReconNet}\cite{shen2019patient} & 29.805±3.124 & 0.660±0.052 & 34.803±5.877 & 0.330±0.051 \\
 & \textit{X2CT-CNN}\cite{ying2019x2ct} & 31.669±3.296 & 0.709±0.052 & 34.948±6.388 & 0.264±0.039 \\
 & \textit{X2CT-GAN}\cite{ying2019x2ct} & 29.878±2.914 & 0.633±0.059 & 29.788±4.835 & 0.308±0.039 \\
 & \textit{MFCT-GAN}\cite{jiang2022mfct} & 29.989±3.392 & 0.641±0.057 & 29.313±5.133 & 0.322±0.045 \\
 & \textit{SdCT-GAN-A} & \textbf{31.979±3.314} & \textbf{0.710±0.053} & 23.714±4.410 & \textbf{0.260±0.038} \\
 & \textit{SdCT-GAN-B} & 31.937±3.350 & 0.708±0.053 & \textbf{22.640±4.380} & 0.262±0.038 \\ 
\bottomrule
\end{tabular*}
\end{table}

\subsubsection{Generator Loss}
We denote the training dataset by $S = \left \{ (x_{i}  , y_{i}  )|1 \le  i \le  N \right \} $ , where $\left ( x_{i}  , y_{i}  \right ) $ is the $i$-th pair of X-rays and CT in $S$, and $N$ is the number of pairs in $S$. Inputting $x_{i}$ to the generator $G\left ( \cdot  \right ) $, we can obtain our reconstructed CT $\hat{y_{i}}  = G(x_{i})$.

CT images need to provide clinical values, so the synthetic CT task should have higher semantic similarity requirements than other natural image synthesis tasks. Hence, we propose a new loss function ${\cal L}_{g3DPcept}$, formulated as follows, to improve the perceptual quality of the reconstructed CT images in SdCT-GAN. Motivated by the Learning Perceptual Image Patch Similarity (LPIPS) \cite{cvpr/ZhangIESW18} as a metric, ${\cal L}_{g3DPcept}$  computes the difference at each layer of $D$ of the CT image cross-section.  For perception, we compute the weighted $\hat{y_{i}}$ and $y_{i}$ distances between two VGG16 feature spaces, where the weights are pre-trained to mimic human perceptual judgments. 

\begin{equation}{\cal L}_{g3DPcept} = \frac{1}{ND} \sum_{i=1}^{N} \sum_{j=1}^{D} {\left \| {Vgg\left (\psi (y _{i})_{j} \right  ) - Vgg\left ( \psi({\hat{y}}_{i})_{j} \right )} \right \| }_{2}^{2}\hspace{0.2em}.\label{eq:G5}\end{equation} where $\psi(\cdot )$ indicates the 3D image is cut from the depth direction to the 2D image of the coronal plane, as in {Figure~\ref{fig: SdCT-GAN-frame}(d).

The overall generator loss ${\cal L}_{G}$ consists of the generation adversarial loss ${\cal L}_{g}$ and generation reconstruction loss ${\cal L}_{gReconstr}$. Generation reconstruction loss ${\cal L}_{gReconstr}$ concludes the reconstruction voxel-wise loss ${\cal L}_{gVoxel}$, the projection map loss ${\cal L}_{gPm}$, and the proposed complementary loss of 3D coronal slices ${\cal L}_{g3DPcept}$. The loss term for each is multiplied with session weight ${\lambda}$ to control their relative contributions. Finally, ${\cal L}_{G}$ is defined as:
\begin{equation}{\cal L}_{G} = {\lambda} _{1}{\cal L}_{g}+{\cal L}_{gReconstr}\hspace{0.2em}.\end{equation}
\begin{equation}{\cal L}_{gReconstr} = {\lambda} _{2}{\cal L}_{gVoxel}+{\lambda}_{3}{\cal L}_{gPm}+{\lambda}_{4}{\cal L}_{g3DPcept} \hspace{0.2em}.\end{equation}
In our following experimental studies, the setting is ${\lambda} _{1}=0.1$, ${\lambda}_{2}={\lambda}_{3}=10$, ${\lambda} _{4}=0.01$. 

${\cal L}_{g}$ is based on $L_2$-norm loss, denoted by ${\left \|\cdot  \right \|} _{2} $, measuring the ability for the generator to fool discriminator, defined as:
\begin{equation}{\cal L}_{g} = \frac{1}{N} \sum_{i=1}^{N} {\left \| \textbf{1}-D\left ( {{\hat{y}}}_{i} \mid x_{i} \right )  \right \| }_{2}^{2} \hspace{0.2em}.\label{eq:G2}\end{equation}
 where \textbf{1} = ${\left [ 1,1,\cdots ,1  \right ]}^{T}  $ is the respective ground-truth with a tensor of integer $a$ that shares the same size as the output of discriminator $D\left ( \cdot  \right ) $. $D\left ( {{\hat{y}}}_{i} \mid x_{i} \right )$ represents the discriminator network which is trained to differentiate between prediction $\hat{y_{i}}$ and the target $y_{i}$ with given input $x_{i}$.

${\cal L}_{gVoxel}$ can reduce the numerical differences between the synthesised 3D image and the real image, which utilizes smooth  $L_2$-norm loss to assess the voxel-wise loss between the output of generator $\hat{y_{i}}$ and the corresponding real CT data $y_{i}$, defined as:
\begin{equation}{\cal L}_{gVoxel} = \frac{1}{N} \sum_{i=1}^{N}{\left \| y _{i} - {\hat{y}}_{i} \right \| }_{2}^{2}\hspace{0.2em}.\label{eq:G3}\end{equation}

${\cal L}_{gPm}$ exploits the $L_1$-norm to calculate the projection loss of the synthetic image in the three orthogonal projection planes (axial, coronal, and sagittal as in {Figure~\ref{fig: SdCT-GAN-frame}(c)}), defined as:

\begin{equation} \begin{aligned}{\cal L}_{gPm} = \frac{1}{3} &[{ \left \|P_{a}(y _{i}) - P_{a}(\hat{y}  _{i}) \right \|}_{1} + \\
 &{ \left \|P_{c}(y _{i}) - P_{c}(\hat{y}  _{i}) \right \|}_{1}  + \\
&{ \left \|P_{s}(y _{i}) - P_{s}(\hat{y}_{i}) \right \|}_{1} 
 ]\hspace{0.2em}.\label{eq:G4}\end{aligned}\end{equation}
where $P_{a}(\cdot)$,$P_{c}(\cdot)$, and $P_{s}(\cdot)$ denote images in the axial, coronal, and sagittal projections, respectively.

\subsection{DAE (Discriminator with AutoEncoder)}
Compared to general image synthesis tasks, medical CT image synthesis requires a model with finer and more comprehensive feature extraction capability due to the low variance of medical image data of the same organ. Many recent studies have enhanced the feature extraction simply by data augmentation\cite{tran2021data,liang2022sketch}, network structure adjustment\cite{ying2019x2ct, ratul2021ccx,jiang2022mfct}, or objective function redesign\cite{jiang2021synthesis}. However, the vital role and influence of the discriminator's indirect gradient on the generator during adversarial training has been disregarded. According to game theory, the generator cannot constantly be improved at the expense of the discriminator. Otherwise, with a limited training dataset, the generator or discriminator may fall into ill-posed situations such as model collapse, leading to a sub-optimal data distribution estimation.

To address this issue, we propose a novel 3D auto-encoder structure for the discriminator, as shown in {Figure~\ref{fig:discrimintor}}. In {Figure~\ref{fig:discrimintor}(a)}, the traditional discriminator component is responsible for learning to differentiate between synthetic data generated by the generator and real data. By extracting comprehensive and essential features, the discriminator can better guide the generator's learning through adversarial training. In {Figure~\ref{fig:discrimintor}(b)}, the newly added autoencoder network structure cleverly drives the extraction of more useful information from the component in {Figure~\ref{fig:discrimintor}(a)}. Specifically, we establish reconstruction constraints between ${I_{part}}$ and ${I^{'}}_{part}$, and between ${I^{'}}$ and ${I^{'}}$. As ${I^{'}}$ and ${I^{'}}_{part}$ are generated by different simple decoders using $f_2$ and cropped $f_1$, respectively, and $f_1$ and $f_2$ are obtained through multi-layer encoding in the component of {Figure~\ref{fig:discrimintor}(a)}, this form of autoencoder training serves as an effective regularization strategy during iterative training. It encourages the discriminator to extract relevant local information from $f_1$ and global information from $f_2$, while also optimizing the decoder in {Figure~\ref{fig:discrimintor}(b)}}.

In summary, there are several novelties in the proposed approach: 
\begin{itemize}
\item It is the first time to apply the autoencoder reconstruction learning discriminator in GAN to 3D images, promoting the network's ability from 2D to 3D processing. And the comparative effects of different reconstruction loss functions in the discriminator on the quality of the synthetic images have been explored.
\item A more elaborate feature map cropping is utilized. We use two decoders on two different scales for the feature maps: $f_{1}$ on $16^{3}$ and $f_{2}$ on $8^{3}$, and cut out half of the height, width, and depth of $f_{1}$ at random, then cut out the same part of the real image to get $I_{part}$, as shown in {Figure~\ref{fig:discrimintor}}. After that, we resize the actual image to obtain $I$. The decoders generate the ${I^{'}}_{part}$ component from the cropped $f_{1}$ and the ${I^{'}}$ part from $f_{2}$, which is only trained on real data.
\item A conditional constraint on the input data X-ray is imposed, combining X-ray and CT and then feeding to the discriminator.
\item The effect of different loss functions on the auto-encoder structure is explored
\end{itemize}

\subsubsection{Discriminator Loss} 
The discriminator loss ${\cal L}_{D}$, defined as follows, is composed of the discrimination adversarial loss ${\cal L}_{d}$ and the reconstruction constraint loss ${\cal L}_{dReconstr}$, and ${\cal L}_{dReconstr}$ carries the reconstruction loss of the simple decoder, which consists of the voxel reconstruction loss function ${\cal{L}}_{dVoxel}$ and the perceptual reconstruction loss function ${\cal{L}}_{d3DPecpt}$ to ensure a balance between the generator and the discriminator. Loss terms are multiplied with weights ${\lambda} _{1}$, ${\lambda} _{2}$ and ${\lambda} _{4}$, which have the same values as the generator.

\begin{equation}{\cal L}_{D} = {\lambda} _{1}{\cal L}_{d}+{\cal L}_{dReconstr}\hspace{0.2em}.\end{equation}
\begin{equation}{\cal L}_{dReconstr} = {\lambda} _{2}{\cal L}_{dVoxel}+{\lambda}_{4}{\cal L}_{d3DPcept} \hspace{0.2em}.\end{equation}

${\cal L}_{d}$ estimates the power of 3DSelf-supervised $D\left ( \cdot  \right ) $ to distinguish between the real and generated data, defined as:
\begin{equation}{\cal L}_{d} = \frac{1}{N} \sum_{i=1}^{N} ({\left \| \textbf{0}-D\left ( {{\hat{y}}}_{i} \mid x_{i} \right )  \right \| }_{2}^{2} + {\left \| \textbf{1}-D\left ( {{y}}_{i} \mid x_{i} \right )  \right \| }_{2}^{2}) \hspace{0.2em}.\label{eq:D2}\end{equation}

${\cal L}_{dVoxel}$ computes the volexl-wise matching differences between $I_{part}$ and ${I^{'}}_{part}$, $I$ and ${I^{'}}$, defined as:

\begin{equation}{\cal L}_{dVoxel} = \frac{1}{N} \sum_{i=1}^{N}{\left \|f - p \right \| }_{2}^{2}\hspace{0.2em}.\label{eq:D3}\end{equation}
where $f$ is the intermediate feature map from simple decoders in the discriminator and $p$ is the real CT image after relevant processing operations. 

${\cal{L}}_{d3DPecpt}$ calculates the perception matching errors between $I_{part}$ and ${I^{'}}_{part}$, $I$ and ${I^{'}}$ to assess the discriminator's capacity to extract features, defined as:
\begin{equation}{\cal{L}}_{d3DPecpt} = \frac{1}{ND} \sum_{i=1}^{N} \sum_{j=1}^{D} {\left \| {Vgg(\psi(f)_{j}) - Vgg(\psi(p)_{j})}\right \| }_{2}^{2}\hspace{0.2em}.\label{eq:D4}\end{equation}

\begin{table}[!t]
\centering
\caption{The impact of different hyperparameters and loss functions on the DAE (Mean ± Standard Deviation). It is worth noting that the DAE structure is the main loss function is auxiliary, the loss function is dependent on the DAE structure is also considered part of the DAE.}\label{tab:DAE_Ablation}
\begin{tabular*}{\textwidth}{@{\extracolsep\fill}lcccccc}
\toprule%
\multicolumn{3}{c}{\textbf{DAE Combination}} &  & \multicolumn{2}{c}{\textbf{Metrics}} \\ \cmidrule{1-3}\cmidrule{5-6}%
Hyperparameters & ${\cal L}_{dVoxel}$ & ${\cal L}_{d3DPcept}$ &  & PSNR & LPIPS \\ \hline
\multirow{3}{*}{\begin{tabular}[c]{@{}c@{}}a$\colon$\\      ${\lambda}_{2}=10$\\      ${\lambda}_{4}=0.01$\end{tabular}} & {\checkmark} &  &  & \textbf{32.248±3.278} & 29.778±6.004 \\
 &  & {\checkmark} &  & 32.195±3.285 & 29.760±5.904 \\
 & {\checkmark} & {\checkmark} &  & 30.525±2.758 & \textbf{26.508±4.225} \\ \hline
\multirow{3}{*}{\begin{tabular}[c]{@{}c@{}}b$\colon$\      ${\lambda}_{2}=100$  \\      ${\lambda}_{4}=0.1$\end{tabular}} & {\checkmark} &  &  & 29.860±2.793 & 29.664±5.032 \\
 &  & {\checkmark} &  & 30.359±2.289 & 27.827±4.568 \\
 & {\checkmark} & {\checkmark} &  & 30.550±3.154 & 28.743±5.346 \\ \hline
\multicolumn{3}{c}{Baseline   X2CT-GAN:${\lambda}_{2}$=10} &  & 29.878±2.914 & 29.788±4.835 \\ \hline
\multicolumn{3}{c}{DAE\_no\_dReconstr} &  & 32.206±3.348 & 32.393±6.174 \\ 
\bottomrule
\end{tabular*}
\end{table}

\section{Experiments}\label{sec4}

In this section, we present the evaluation results of the proposed SdCT-GAN model with conventional metrics such as PSNR (Peak Signal-to-Noise Ratio), SSIM (Structural SIMilarity)\cite{sara2019image}, and NRMSE (Normalized Root-Mean-Square Error)\cite{shcherbakov2013survey}. In addition, as elaborated in {Figure~\ref{fig:first_imag}}, PSNR and SSIM may not be a good fit for quality measures of internal details, we apply a new quantitative metric of LPIPS (Learned Perceptual Image Patch Similarity) to assess the internal details of 3D images, which utilizes a pre-trained VGG16 network to gauge perceptual differences between 128 cross-sectional images of CT scans (as shown {Figure~\ref{fig: SdCT-GAN-frame}(d)}). And smaller differences indicate better performance. 

To demonstrate the efficacy and validity of our proposed approach, we have reproduced the current state-of-the-art models such as 2DCNN\cite{henzler2018single}, ReconNet\cite{shen2019patient}, X2CT-CNN\cite{ying2019x2ct}, X2CT-GAN\cite{ying2019x2ct}, and MFCT-GAN\cite{jiang2022mfct} (not open-sourced), and have conducted comparative empirical studies to exhibit the outperformance of our proposed method over these baselines. Here we rule out CCX-RAYNET\cite{ratul2021ccx} (not open-sourced) for comparison because it is not an end-to-end network structure, relying heavily on additional input from a prior semantic segmentation result.

Moreover, in order to showcase the scalability of our proposed DAE (Discriminator with AutoEncoder ) architecture, we have applied it to a task of reconstructing three-dimensional to three-dimensional (3D-3D: PD-T2) MRI images, which allows us to conduct a comparative analysis with Ea-GANs \cite{yu2019ea}.

\begin{table}[!t]
\caption{Evaluation of different components (Mean ± Standard Deviation). The results in the first row indicate the use of X2CT-GAN as the baseline model. The discriminator network structure, referred to as DAE, is depicted in {Figure~\ref{fig:discrimintor}}. DAE-A specifically employs ${\cal L}_{{Voxel}}$ as the only component in the discriminator's reconstruction loss function, whereas DAE-B incorporates both ${\cal L}_{{dVoxel}}$ (voxel reconstruction loss function) and ${\cal L}_{{d3DPecpt}}$ (perceptual reconstruction loss function) in the discriminator's reconstruction loss function. Additionally, the term ${\cal L}_{{g3DPcept}}$ serves as a constraint in generator, denoted as g3DPcept.}\label{tab:ablation_res} 
\begin{tabular*}{\textwidth}{@{\extracolsep\fill}lccccccc}
\toprule%
\multicolumn{4}{c}{\textbf{Combination}} & \textbf{} & \multicolumn{2}{c}{\textbf{Metrics}} \\ \\\cmidrule{1-4}\cmidrule{6-7}%
\textbf{DAE-A} & \textbf{DAE-B} & \textbf{g3DPcept} & \textbf{SGG} & \textbf{} & \textbf{RSNR} & \textbf{LPIPS} \\
\midrule
 &  &  &  &  & 29.878±2.914 & 29.788±4.835 \\
{\checkmark} &  &  &  &  & 32.248±3.278 & 29.778±6.004 \\
 & {\checkmark} &  &  &  & 30.525±2.758 & 26.508±4.225 \\
 &  & {\checkmark} &  &  & 31.068±3.363 & 24.565±4.292 \\
 &  &  & {\checkmark} &  & 30.623±3.133 & 28.816±4.819 \\
{\checkmark} &  & {\checkmark} &  &  & 31.880±3.352 & 24.173±4.727 \\
{\checkmark} &  & {\checkmark} & {\checkmark} &  & 31.979±3.314 & 23.714±4.409 \\
 & {\checkmark} & {\checkmark} &  &  & 31.863±3.332 & 23.177±4.435 \\
 & {\checkmark} &  & {\checkmark} &  & 31.945±3.207 & 28.699±5.332 \\
 &  & {\checkmark} & {\checkmark} &  & 31.274±3.260 & 24.237±4.174 \\
 & {\checkmark} & {\checkmark} & {\checkmark} &  & 31.937±3.350 & \textbf{22.640±4.38} \\
\botrule
\end{tabular*}
\end{table}

\begin{table}[!t]
\caption{Comparing the effects of different sobel filtering methods on SdCT-GAN-B.}\label{tab:otherSobel}
\begin{tabular*}{\textwidth}{@{\extracolsep\fill}lccccccccc}
\toprule%
\multicolumn{6}{c}{Combination} &  & \multicolumn{2}{c}{Metrics} \\\cmidrule{1-6}\cmidrule{8-9}%
DAE-B & g3DPcept & SGG & gEa & dEa & sWeight &  & RSNR & LPIPS \\
\midrule
& {\checkmark} &  &  &  &  &  & 31.068±3.363 & {\color[HTML]{68CBD0} 24.565±4.292} \\
 & {\checkmark} &  & {\checkmark} &  &  &  & 31.237±3.176 & {\color[HTML]{68CBD0} 25.829±4.465} \\
{\checkmark} & {\checkmark} & {\checkmark} &  &  &  &  & 31.937±3.350 & \textbf{22.640±4.380} \\
{\checkmark} & {\checkmark} &  &  & {\checkmark} &  &  & 32.017±3.352 & 23.782±4.343 \\
{\checkmark} & {\checkmark} &  &  &  & {\checkmark} &  & 31.170±3.216 & 30.610±5.380 \\
{\checkmark} & {\checkmark} & {\checkmark} &  & {\checkmark} &  &  & 31.958±3.345 & 22.846±4.298 \\
\botrule
\end{tabular*}

\end{table}

\subsection{Dataset}
\subsubsection{LIDC-IDRI}
The benchmark in our experiments is a biplanar X-Rays and CT-paired dataset, that comes from Ying et al\cite{ying2019x2ct} where synthetic X-rays from real CT scans by digitally reconstructed radiograph (DRR) with CycleGAN\cite{caponetti19903d} is provided. And the 1018 chest CT scans are from the LIDC-IDRI dataset\cite{choy20163d}. In our settings, 916 and 102 CT scans are for training and testing, respectively. 
\subsubsection{IXI}
The IXI dataset comprises non-intracranially stripped magnetic resonance (MR) images of the brain, encompassing 578 subjects across five modalities: T1, T2, PD, MRA, and DTI. In our study, we followed a processing protocol similar to Ea-GANs \cite{yu2019ea}, whereby the images were standardized to a uniform size of 128 × 128 × 128 and the raw intensity values were scaled to the range of [-1, 1]. To ensure unbiased evaluation, we employed a five-fold cross-validation strategy, where the dataset was partitioned into training and test sets with 4/5 and 1/5 of the subject samples, respectively.

\subsection{Implementation Details}
Our model is implemented by Python 3.7 PyTorch on a workstation with NVIDIA-Tesla-V100-SXM2-32GB, in which the generator and discriminator are trained alternatively with the standard process\cite{mirza2014generative}. The Adam 
optimizer\cite{kingma2014method} is used to train our network, and the initial learning rate is 2e-4. After training for 50 epochs, we apply a linear learning rate decay policy to decrease it to 0. And we train our model for a total of 100 epochs, during which we apply instance normalization instead of batch normalization. Limited by the GPU memory, the batch size is set to two in our settings. Moreover, in the comparative experiments, we keep the hyperparameters consistent with the baseline of X2CT-GAN\cite{ying2019x2ct} except for the weight coefficients of the loss function ${\cal L}_{d3DPcept}$ that we have newly introduced in the proposed model.

In the scalability experiments of DAE, we have the same training method and hyperparameter settings as Ea-GANs \cite{yu2019ea}, the only difference is the addition of the hyperparameters for the reconstruction loss in DAE and the specific hyperparameter settings are as follows: in DAE-gEaGAN, ${\lambda}_{2}$=10 and ${\lambda}_{4}$=0.02, while in DAE-dEaGAN, ${\lambda}_{4}$=200 and ${\lambda}_{4}$=0.2.

\begin{figure*}[!t]
\centering
\includegraphics[width=0.9\textwidth]{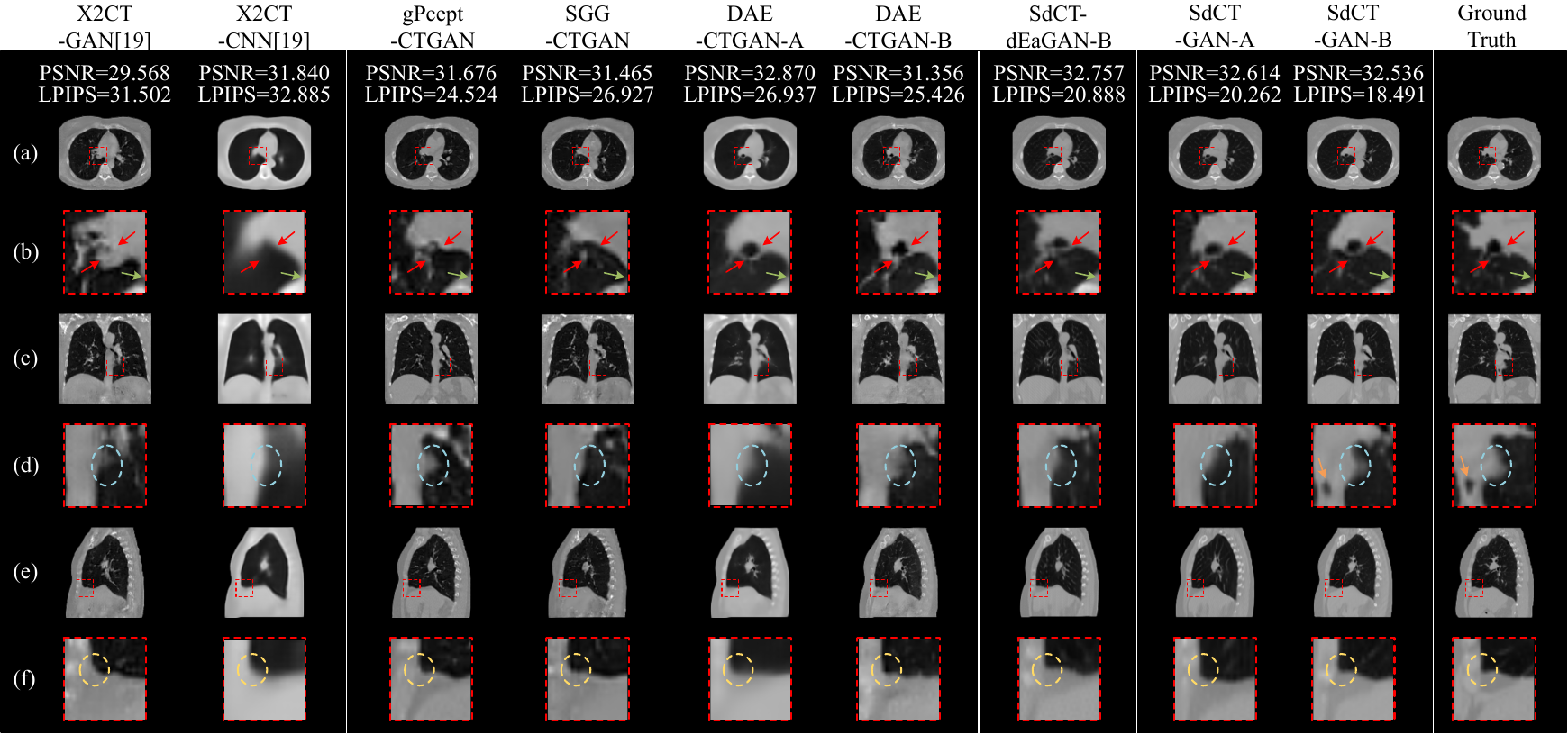}
\caption{Visualization of synthesized image details in the ablation experiments of SdCT-GAN.(a) axial slices, (b) zoomed parts of axial slices, (c) coronal slices, (d) zoomed parts of coronal slices, and (e) sagittal slices, (f) zoomed parts of sagittal slices. SdCT-dEaGAN-B denotes the modification made to our model SdCT-GAN-B by removing the SGG module and incorporating the dEa module. The dEa module involves adding Sobel filtered gradients as guidance in the objective function and inputting the Sobel feature maps to the discriminator of DAE.}\label{fig:CT-detail-ablation}
\end{figure*}

\subsection{Qualitative Results}
We first present the qualitative CT reconstruction visualization in {Figure~\ref{fig:reconstructed_CT}}. It can be clearly seen that the lung texture is finer and the noises are significantly reduced in the 3D lung rendering map of our reconstructed CT image compared with other baselines; similarly, the 3D stereogram of our reconstructed CT image has a more precise edge contour, especially at the locations marked by the red and blue boxes.

{Figure~\ref{fig:error_map}} illustrates the comparisons of error maps between the synthesized CT and the corresponding ground truth CT in the transverse, sagittal, and coronal planes. It is evident from the images that our SdCT-GAN produces distributions that are most similar to the ground truth data. By contrast, the 2DCNN method exhibits a large amount of bright green and some red colors, indicating significant discrepancies between the synthesized and real images. Although the X2CT-CNN method mostly displays light blue colors and fewer bright green colors compared to X2CT-GAN and MFCT-GAN, it presents distinct and varying patterns of textures in the error maps, indicating over-smoothing of the generated images.

In {Figure~\ref{fig:detail_map}}, it can be observed that our model SdCT-GAN-B performs the best in the reconstruction of the internal details, structural contours and textures of the 3D images compared to other methods, as highlighted by the red boxes of the boundaries between the lung tissue and blood vessels, ascending aorta, trachea, right pulmonary artery branches, vertebral arch and descending aorta. And, the results of {Figure~\ref{fig:detail_map}} are consistent with those of {Figure~\ref{fig:error_map}}, where X2CT-CNN is indeed too smooth.

\begin{figure}[!t]
\centering
\includegraphics[width=0.9\textwidth]{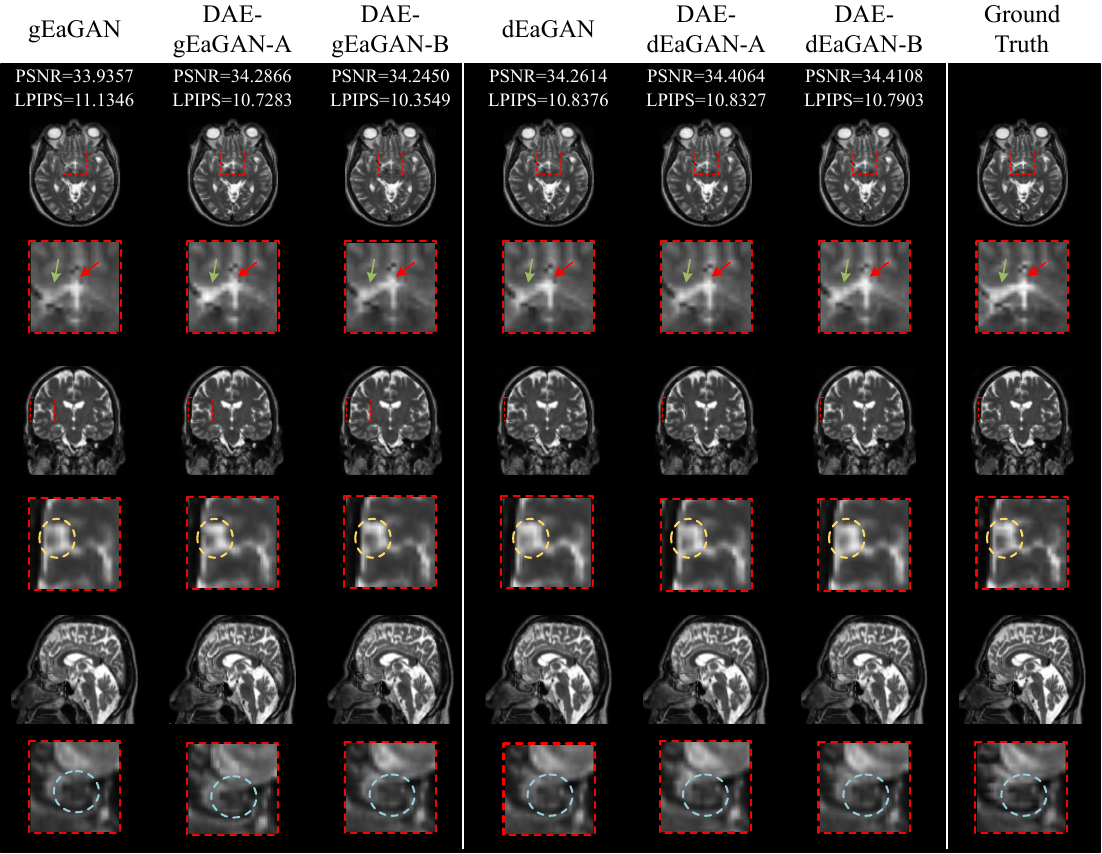}
\caption{Visualization of synthesized image details in the ablation experiments of SdCT-GAN.(a) axial slices, (b) zoomed parts of axial slices, (c) coronal slices, (d) zoomed parts of coronal slices, and (e) sagittal slices, (f) zoomed parts of sagittal slices.}\label{fig:mri_result_detail}
\end{figure}

\subsection{Quantitative Results}
{Table \ref{tab:quantitative_res}} provides a quantitative summary of the empirical evaluations. As 
for the baselines, we reproduce them as much as possible to demonstrate the most performance. And the final results show that our proposed SdCT-GAN performs the best in PSNR, SSIM, LPIPS, and NRMSE for both single and dual views. Compared to baseline X2CT-GAN \cite{ying2019x2ct}, our SdCT-GAN-B shows an overall improvement of 3.025\% in PSNR on a single view; a significant improvement of 8.015\% in SSIM, an overall decrease of 5.77\% in LPIPS, and an overall decrease of 4.166\% in NRMSE (Hu) values; on multi-perspectives, the overall improvement in PSNR values is 6.891\% with the substantial improvement of 11.848\% in SSIM, and the general reduction in LPIPS and NRMSE (Hu) values is 23.996\% and  14.935\%, respectively. 

Furthermore, SdCT-GAN-A, which adopts the MSE loss function only in the discriminator, leads to slightly better pixel-level reconstruction performance compared to SdCT-GAN-B, both in single-view and multi-view scenarios. This observation is further supported in {Figure~\ref{fig:error_map}}. While both methods exhibit similar performance in terms of PSNR and SSIM metrics, SdCT-GAN-B demonstrates a distinct advantage over SdCT-GAN-A in terms of the LPIPS metric, both in single-view and multi-view settings. LPIPS reflects the differences in visual perception, where a smaller LPIPS value indicates better restoration of overall structure and fine textures. This observation is further supported by the results depicted in Figures~\ref{fig:reconstructed_CT}, \ref{fig:error_map}, and \ref{fig:detail_map}, where SdCT-GAN-B achieves the lowest LPIPS score, indicating its superior performance in reconstructing the overall shape and content details of the images. 


\begin{table}[!t]
\centering
\renewcommand\arraystretch{1.2}
\caption{The effect of DAE applied to gEaGAN and dEaGAN\cite{yu2019ea} (Mean ± Standard Deviation).dEaGAN is the integration of edge information into adversarial learning based on gEaGAN.}\label{tab:extendMRI}
\begin{tabular*}{\textwidth}{@{\extracolsep\fill}lccccc}
\toprule
\multirow{2}{*}{\textbf{Method}} & \multicolumn{4}{c}{\textbf{Metrics}} \\
\cmidrule(lr){2-5}
 & \textbf{RSNR} & \textbf{SSIM} & \textbf{LPIPS} & \textbf{NRMSE(Hu)} \\ 
\midrule
gEaGAN & 33.623±2.001 & 0.9302±0.020 & 11.750±2.712 & 0.0706±0.020 \\
DAE-gEaGAN-A & 33.703±2.017 & 0.9315±0.020 & 11.538±2.705 & 0.0700±0.021 \\
DAE-gEaGAN-B & 33.731±2.048 & 0.9317±0.020 & \textbf{11.153±2.658} & 0.0696±0.021 \\
dEaGAN & 33.746±2.064 & 0.9316±0.021 & 11.517±2.678 & 0.0697±0.021 \\
DAE-dEaGAN-A & 33.841±2.005 & 0.9322±0.020 & 11.485±2.664 & 0.0686±0.021 \\
DAE-dEaGAN-B & 33.784±2.719 & 0.9317±0.021 & 11.484±2.744 & 0.0692±0.021 \\ 
\bottomrule
\end{tabular*}
\end{table}

\subsection{Ablation Study}
The conceptualization and establishment of hyperparameters in our method follow an analytical and progressive approach. Initially, we need to determine the range of hyperparameters (weights of voxel reconstruction loss and perceptual loss, denoted as ${\lambda}_{2}$ and ${\lambda}_{4}$, respectively) in the Discriminator-with-AutoEncoder (DAE). This is a crucial prerequisite for exploring the effectiveness of the model. Therefore, we have formulated two principles in setting the hyperparameters, which can greatly assist in the applications of DAE in other 3D GAN network models. \begin{itemize}
\item Principle 1: The generator in GAN networks typically includes voxel reconstruction loss. Consequently, we consider that the values of ${\cal L}_{gVoxel}$ and ${\cal L}_{dVoxel}$ at epoch=0 should either be at the same order of magnitude, or their difference in the order of magnitude be gradually increasing.
\item Principle 2: On the basis of Principle 1. The values of ${\cal L}_{dVoxel}$ and ${\cal L}_{d3DPcept}$ at epoch=0 should be within the same order of magnitude in the first place.
\end{itemize}
In {Table~\ref{tab:DAE_Ablation}}, the data in group `a' initially follows Principle 1. We can observe that when ${\lambda}_{2}$=10, the values of ${\cal L}_{gVoxel}$ in the generator and ${\cal L}_{dVoxel}$ in the DAE at epoch=0 are within the same order of magnitude. Subsequently, following Principle 2, we can find that when ${\lambda}_{4}$=0.01, the values of ${\cal L}_{dVoxel}$ and ${\cal L}_{d3DPcept}$ are within the same order of magnitude. 

The data in group `b' in {Table~\ref{tab:DAE_Ablation}} represents an alternative scenario following Principle 1. Next, we identify the optimal values (bolded) for each metric in {Table~\ref{tab:DAE_Ablation}}, which are all in group `a'. Based on this, we have determined the selection of the two hyperparameters in {Table~\ref{tab:DAE_Ablation}}-a. Further analysis of the data in group `a' in {Table~\ref{tab:DAE_Ablation}} reveals that the differences between using ${\cal L}_{dVoxel}$ or ${\cal L}_{d3DPcept}$ as the sole reconstruction loss function in the DAE are not significant. When applied individually, ${\cal L}_{dVoxel}$ performs slightly better than ${\cal L}_{d3DPcept}$ in terms of PSNR. However, when both loss functions are employed, there is a decrease of 5.34\%, in PSNR, respectively, while LPIPS shows a 13.36\% improvement. 

To further investigate the impact of different reconstruction loss functions on the effectiveness of the model, we have selected the first and third variants from group `a' in {Table~\ref{tab:DAE_Ablation}}, denoted as DAE-A and DAE-B, respectively, for subsequent ablation experiments.

We further conducted ablation studies to evaluate the effectiveness of different components in our proposed SdCT-GAN. According to the results in {Table~\ref{tab:ablation_res}}, each component individually has demonstrated positive improvement compared to the baseline model X2CT-GAN. Interestingly, when using the DAE-A alone, there is a significant improvement in PSNR (7.932\%), compared to the overall X2CT-GAN baseline, but the improvement in LPIPS has been little. Based on the quantitative and qualitative results, it is evident that the model performs well in three metrics but presents limitations in LPIPS, possibly due to the generated images resembling smoother X2CT-CNN results. This occurs because the generator needs to be well incentivized to focus on the overall and local information while the driving power from the discriminator is not enough. Therefore, when the generator is equipped with the proposed supplementary learning of perceptual image patch similarity module, denoted as ${\cal L}_{g3DPcept}$, the model is greatly empowered and the performance can be boosted significantly,  either in combination with DAE-A or DAE-B.  

Specifically, with the addition of ${\cal L}_{g3DPcept}$ into DAE-A, compared to using ${\cal L}_{g3DPcept}$ alone, there has been  2.613\% increase in PSNR and 1.596\% decrease in LPIPS. Compared to the X2CT-GAN baseline, we can see 6.701\% increase in PSNR and 18.850\% decrease in LPIPS. Similarly, with the addition of ${\cal L}_{g3DPcept}$ into DAE-B, we can observe 2.559\% increase in PSNR and 5.650\% decrease in LPIPS. Compared to the X2CT-GAN baseline, there has been 6.644\% increase in PSNR and 22.194\% decrease in LPIPS. Lastly, the introduction of the SGG module can also lead to consistent improvements in the model's performance, regardless of whether DAE-A or DAE-B is used. The SGG module, whether used alone or combined with any component, consistently enhances the model's performance. Based on these results, we can conclude that DAE-B exhibits a clear advantage in terms of LPIPS, while DAE-A demonstrates superiority in terms of PSNR. It is also confirmed in Qualitative Result that the SdCT-GAN-B generated clearer images than the SdCT-GAN-A method.

Furthermore, we have conducted additional ablation experiments as shown in {Table~\ref{tab:otherSobel}}. The results suggest that, with the incorporation of our proposed DAE and ${\cal L}_{g3DPcept}$ modules, our SGG module is better propelled compared to other Sobel filtering methods. Moreover, we can observe that adding both ${\cal L}_{g3DPcept}$ and SobelLoss (from gEa) in the loss function can lead to underperformance, as indicated by the blue font in {Table~\ref{tab:otherSobel}}. That is to say, the inclusion of ${\cal L}_{g3DPcept}$ in gEa may cause an increase in the LPIPS metric, and since dEa also contains gEa, it has a certain impact on the LPIPS metric as well. By contrast, our SGG module can consistently improve the LPIPS metric with the fusion of the proposed DAE and ${\cal L}_{g3DPcept}$.

The results shown in {Figure~\ref{fig:CT-detail-ablation}} align with the aforementioned analysis. Firstly, it is evident from the images that the DAE-CTGAN-A and X2CT-CNN method produces blurred edges and smoother images compared to X2CT-GAN, which indicates the preservation of subtle contour information and is consistent with the lower LPIPS scores associated with better preservation of fine structures. Furthermore, in {Table~\ref{tab:otherSobel}}, while SdCT-dEaGAN-B and our proposed SdCT-GAN-A and SdCT-GAN-B methods exhibit comparable performance in terms of the PSNR metric, SdCT-GAN-B achieves the lowest LPIPS score. The results in {Figure~\ref{fig:CT-detail-ablation}} further justify that SdCT-GAN-B approximates the real images the best, particularly in terms of image detail restoration (highlighted by yellow arrows in the fourth row). Therefore, it can be concluded that when there is little difference in PSNR, LPIPS is able to serve as a more pragmatic and complementary metric for evaluating the quality of synthesized images.

\subsection{Scalability Study}
In the scalability experiments, the two principles mentioned above still apply. The results in Table \ref{tab:extendMRI} show that both gEaGAN and dEaGAN equipped with the DAE discriminator have led to a stable improvement in the model, especially DAE-gEaGAN-B can significantly outperform dEaGAN in the metric of LPIPS, with a 3.161\% decrease. However, from the results of DAE-dEaGAN-A and DAE-dEaGAN-B in {Table~\ref{tab:extendMRI}}, we have found that the improvement effect of DAE applied to the upper dEaGAN is not significant, and we believe that the additional integration of a sobel feature map of dEaGAN changes the data distribution of the GAN, so the use of the original tuning method is not helpful enough. In {Table~\ref{fig:mri_result_detail}}, the results show that the images synthesized by gEaGAN-B are closer to the distribution of real images, as indicated by the annotated features, where a hollow `Z' can be observed in the blue circle of the last row.

\section{Conclusion}
This paper proposes a novel self-supervised generative adversarial network structure of SdCT-GAN, which combines effective regularization strategies, conditional constraints, and a Sobel Gradient Guider (SGG) module to address the challenging task of reconstructing 3D CT images using only two orthogonal 2D X-ray projections. Specifically, the model demonstrates significant advantages in preserving voxel similarity, edge similarity, and perceptual similarity of global and local features during the synthesis process, surpassing various state-of-the-art X-ray-based CT image synthesis methods in preserving details, textures, and structural information. Additionally, the introduced discriminator structure exhibits remarkable versatility in MRI image synthesis tasks. Furthermore, we experimentally validate the limitations of image similarity evaluations solely based on voxel-level measurements such as PSNR, SSIM, and NRMSE, and propose to apply the LPIPS evaluation metric, which directly quantifies the reconstruction quality of internal texture and structural details, providing a more comprehensive and pragmatic evaluation framework for image synthesis tasks. 

The objective of this work is not to replace CT with X-rays but to assist in diagnosis. While the proposed method can accurately reconstruct overall structures and lung texture details to some extent, it is still unable to capture fine-grained lung nodules. However, the proposed method may still find suitable applications in clinical practices. For example, it can accurately measure the size of major organs such as the lungs, hearts, and livers, or diagnose poorly positioned organs on the reconstructed CT scans. Moreover, it can also be used for dose planning in radiation therapy, preoperative planning and intraoperative guidance in minimally invasive interventions. Generally speaking, the proposed approach could serve as a valuable enhancement for low-cost X-ray machines, as doctors can obtain CT-like 3D volumes with good precision. 

In the future, we will conduct more systematical statistical experiments to combine the proposed method with X-ray a priori knowledge and clinical studies to further identify and quantify the potential of the proposed method.

\section*{Code and data availability}
The source code and data of our work are available at \url{https://github.com/csqvictory/SdCT-GAN}


\bibliography{sn-article}

\begin{thebibliography}{53}
\ifx \bisbn   \undefined \def \bisbn  #1{ISBN #1}\fi
\ifx \binits  \undefined \def \binits#1{#1}\fi
\ifx \bauthor  \undefined \def \bauthor#1{#1}\fi
\ifx \batitle  \undefined \def \batitle#1{#1}\fi
\ifx \bjtitle  \undefined \def \bjtitle#1{#1}\fi
\ifx \bvolume  \undefined \def \bvolume#1{\textbf{#1}}\fi
\ifx \byear  \undefined \def \byear#1{#1}\fi
\ifx \bissue  \undefined \def \bissue#1{#1}\fi
\ifx \bfpage  \undefined \def \bfpage#1{#1}\fi
\ifx \blpage  \undefined \def \blpage #1{#1}\fi
\ifx \burl  \undefined \def \burl#1{\textsf{#1}}\fi
\ifx \doiurl  \undefined \def \doiurl#1{\url{https://doi.org/#1}}\fi
\ifx \betal  \undefined \def \betal{\textit{et al.}}\fi
\ifx \binstitute  \undefined \def \binstitute#1{#1}\fi
\ifx \binstitutionaled  \undefined \def \binstitutionaled#1{#1}\fi
\ifx \bctitle  \undefined \def \bctitle#1{#1}\fi
\ifx \beditor  \undefined \def \beditor#1{#1}\fi
\ifx \bpublisher  \undefined \def \bpublisher#1{#1}\fi
\ifx \bbtitle  \undefined \def \bbtitle#1{#1}\fi
\ifx \bedition  \undefined \def \bedition#1{#1}\fi
\ifx \bseriesno  \undefined \def \bseriesno#1{#1}\fi
\ifx \blocation  \undefined \def \blocation#1{#1}\fi
\ifx \bsertitle  \undefined \def \bsertitle#1{#1}\fi
\ifx \bsnm \undefined \def \bsnm#1{#1}\fi
\ifx \bsuffix \undefined \def \bsuffix#1{#1}\fi
\ifx \bparticle \undefined \def \bparticle#1{#1}\fi
\ifx \barticle \undefined \def \barticle#1{#1}\fi
\bibcommenthead
\ifx \bconfdate \undefined \def \bconfdate #1{#1}\fi
\ifx \botherref \undefined \def \botherref #1{#1}\fi
\ifx \url \undefined \def \url#1{\textsf{#1}}\fi
\ifx \bchapter \undefined \def \bchapter#1{#1}\fi
\ifx \bbook \undefined \def \bbook#1{#1}\fi
\ifx \bcomment \undefined \def \bcomment#1{#1}\fi
\ifx \oauthor \undefined \def \oauthor#1{#1}\fi
\ifx \citeauthoryear \undefined \def \citeauthoryear#1{#1}\fi
\ifx \endbibitem  \undefined \def \endbibitem {}\fi
\ifx \bconflocation  \undefined \def \bconflocation#1{#1}\fi
\ifx \arxivurl  \undefined \def \arxivurl#1{\textsf{#1}}\fi
\csname PreBibitemsHook\endcsname

\bibitem[\protect\citeauthoryear{Taguchi and
  Iwanczyk}{2013}]{taguchi2013vision}
\begin{barticle}
\bauthor{\bsnm{Taguchi}, \binits{K.}},
\bauthor{\bsnm{Iwanczyk}, \binits{J.S.}}:
\batitle{Vision 20/20: Single photon counting x-ray detectors in medical
  imaging}.
\bjtitle{Medical physics}
\bvolume{40}(\bissue{10}),
\bfpage{100901}
(\byear{2013})
\end{barticle}
\endbibitem

\bibitem[\protect\citeauthoryear{Hounsfield}{1995}]{hounsfield1995computerized}
\begin{barticle}
\bauthor{\bsnm{Hounsfield}, \binits{G.N.}}:
\batitle{Computerized transverse axial scanning (tomography): Part i.
  description of system. 1973.}
\bjtitle{The British journal of radiology}
\bvolume{68}(\bissue{815}),
\bfpage{166}--\blpage{72}
(\byear{1995})
\end{barticle}
\endbibitem

\bibitem[\protect\citeauthoryear{Gordon et~al.}{1970}]{gordon1970algebraic}
\begin{barticle}
\bauthor{\bsnm{Gordon}, \binits{R.}},
\bauthor{\bsnm{Bender}, \binits{R.}},
\bauthor{\bsnm{Herman}, \binits{G.T.}}:
\batitle{Algebraic reconstruction techniques (art) for three-dimensional
  electron microscopy and x-ray photography}.
\bjtitle{Journal of theoretical Biology}
\bvolume{29}(\bissue{3}),
\bfpage{471}--\blpage{481}
(\byear{1970})
\end{barticle}
\endbibitem

\bibitem[\protect\citeauthoryear{Herman}{2009}]{herman2009fundamentals}
\begin{bbook}
\bauthor{\bsnm{Herman}, \binits{G.T.}}:
\bbtitle{Fundamentals of Computerized Tomography: Image Reconstruction from
  Projections}.
\bpublisher{Springer},
\blocation{New York}
(\byear{2009})
\end{bbook}
\endbibitem

\bibitem[\protect\citeauthoryear{Gilbert}{1972}]{gilbert1972iterative}
\begin{barticle}
\bauthor{\bsnm{Gilbert}, \binits{P.}}:
\batitle{Iterative methods for the three-dimensional reconstruction of an
  object from projections}.
\bjtitle{Journal of theoretical biology}
\bvolume{36}(\bissue{1}),
\bfpage{105}--\blpage{117}
(\byear{1972})
\end{barticle}
\endbibitem

\bibitem[\protect\citeauthoryear{Beister et~al.}{2012}]{beister2012iterative}
\begin{barticle}
\bauthor{\bsnm{Beister}, \binits{M.}},
\bauthor{\bsnm{Kolditz}, \binits{D.}},
\bauthor{\bsnm{Kalender}, \binits{W.A.}}:
\batitle{Iterative reconstruction methods in x-ray ct}.
\bjtitle{Physica medica}
\bvolume{28}(\bissue{2}),
\bfpage{94}--\blpage{108}
(\byear{2012})
\end{barticle}
\endbibitem

\bibitem[\protect\citeauthoryear{Padole et~al.}{2015}]{padole2015ct}
\begin{barticle}
\bauthor{\bsnm{Padole}, \binits{A.}},
\bauthor{\bsnm{Ali~Khawaja}, \binits{R.D.}},
\bauthor{\bsnm{Kalra}, \binits{M.K.}},
\bauthor{\bsnm{Singh}, \binits{S.}}, \betal:
\batitle{Ct radiation dose and iterative reconstruction techniques}.
\bjtitle{AJR Am J Roentgenol}
\bvolume{204}(\bissue{4}),
\bfpage{384}--\blpage{392}
(\byear{2015})
\end{barticle}
\endbibitem

\bibitem[\protect\citeauthoryear{Sidky and Pan}{2008}]{sidky2008image}
\begin{barticle}
\bauthor{\bsnm{Sidky}, \binits{E.Y.}},
\bauthor{\bsnm{Pan}, \binits{X.}}:
\batitle{Image reconstruction in circular cone-beam computed tomography by
  constrained, total-variation minimization}.
\bjtitle{Physics in Medicine \& Biology}
\bvolume{53}(\bissue{17}),
\bfpage{4777}
(\byear{2008})
\end{barticle}
\endbibitem

\bibitem[\protect\citeauthoryear{Willemink and
  No{\"e}l}{2019}]{willemink2019evolution}
\begin{barticle}
\bauthor{\bsnm{Willemink}, \binits{M.J.}},
\bauthor{\bsnm{No{\"e}l}, \binits{P.B.}}:
\batitle{The evolution of image reconstruction for ct—from filtered back
  projection to artificial intelligence}.
\bjtitle{European radiology}
\bvolume{29},
\bfpage{2185}--\blpage{2195}
(\byear{2019})
\end{barticle}
\endbibitem

\bibitem[\protect\citeauthoryear{Geraldo et~al.}{2016}]{geraldo2016low}
\begin{barticle}
\bauthor{\bsnm{Geraldo}, \binits{R.J.}},
\bauthor{\bsnm{Cura}, \binits{L.M.}},
\bauthor{\bsnm{Cruvinel}, \binits{P.E.}},
\bauthor{\bsnm{Mascarenhas}, \binits{N.D.}}:
\batitle{Low dose ct filtering in the image domain using map algorithms}.
\bjtitle{IEEE Transactions on Radiation and Plasma Medical Sciences}
\bvolume{1}(\bissue{1}),
\bfpage{56}--\blpage{67}
(\byear{2016})
\end{barticle}
\endbibitem

\bibitem[\protect\citeauthoryear{Yuan et~al.}{2018}]{yuan2018adaptive}
\begin{barticle}
\bauthor{\bsnm{Yuan}, \binits{Y.}},
\bauthor{\bsnm{Zhang}, \binits{Y.}},
\bauthor{\bsnm{Yu}, \binits{H.}}:
\batitle{Adaptive non-local means method for denoising basis material images
  from dual-energy ct}.
\bjtitle{Journal of computer assisted tomography}
\bvolume{42}(\bissue{6}),
\bfpage{972}
(\byear{2018})
\end{barticle}
\endbibitem

\bibitem[\protect\citeauthoryear{Dong et~al.}{2019}]{dong2019sinogram}
\begin{bchapter}
\bauthor{\bsnm{Dong}, \binits{X.}},
\bauthor{\bsnm{Vekhande}, \binits{S.}},
\bauthor{\bsnm{Cao}, \binits{G.}}:
\bctitle{Sinogram interpolation for sparse-view micro-ct with deep learning
  neural network}.
In: \bbtitle{Medical Imaging 2019: Physics of Medical Imaging},
vol. \bseriesno{10948},
pp. \bfpage{692}--\blpage{698}
(\byear{2019}).
\bcomment{SPIE}
\end{bchapter}
\endbibitem

\bibitem[\protect\citeauthoryear{Anirudh et~al.}{2018}]{anirudh2018lose}
\begin{bchapter}
\bauthor{\bsnm{Anirudh}, \binits{R.}},
\bauthor{\bsnm{Kim}, \binits{H.}},
\bauthor{\bsnm{Thiagarajan}, \binits{J.J.}},
\bauthor{\bsnm{Mohan}, \binits{K.A.}},
\bauthor{\bsnm{Champley}, \binits{K.}},
\bauthor{\bsnm{Bremer}, \binits{T.}}:
\bctitle{Lose the views: Limited angle ct reconstruction via implicit sinogram
  completion}.
In: \bbtitle{Proceedings of the IEEE Conference on Computer Vision and Pattern
  Recognition},
pp. \bfpage{6343}--\blpage{6352}
(\byear{2018})
\end{bchapter}
\endbibitem

\bibitem[\protect\citeauthoryear{Pan et~al.}{2022}]{pan2022multi}
\begin{barticle}
\bauthor{\bsnm{Pan}, \binits{J.}},
\bauthor{\bsnm{Zhang}, \binits{H.}},
\bauthor{\bsnm{Wu}, \binits{W.}},
\bauthor{\bsnm{Gao}, \binits{Z.}},
\bauthor{\bsnm{Wu}, \binits{W.}}:
\batitle{Multi-domain integrative swin transformer network for sparse-view
  tomographic reconstruction}.
\bjtitle{Patterns}
\bvolume{3}(\bissue{6}),
\bfpage{100498}
(\byear{2022})
\end{barticle}
\endbibitem

\bibitem[\protect\citeauthoryear{Mizusawa et~al.}{2021}]{mizusawa2021computed}
\begin{barticle}
\bauthor{\bsnm{Mizusawa}, \binits{S.}},
\bauthor{\bsnm{Sei}, \binits{Y.}},
\bauthor{\bsnm{Orihara}, \binits{R.}},
\bauthor{\bsnm{Ohsuga}, \binits{A.}}:
\batitle{Computed tomography image reconstruction using stacked u-net}.
\bjtitle{Computerized Medical Imaging and Graphics}
\bvolume{90},
\bfpage{101920}
(\byear{2021})
\end{barticle}
\endbibitem

\bibitem[\protect\citeauthoryear{Zang et~al.}{2021}]{zang2021intratomo}
\begin{bchapter}
\bauthor{\bsnm{Zang}, \binits{G.}},
\bauthor{\bsnm{Idoughi}, \binits{R.}},
\bauthor{\bsnm{Li}, \binits{R.}},
\bauthor{\bsnm{Wonka}, \binits{P.}},
\bauthor{\bsnm{Heidrich}, \binits{W.}}:
\bctitle{Intratomo: self-supervised learning-based tomography via sinogram
  synthesis and prediction}.
In: \bbtitle{Proceedings of the IEEE/CVF International Conference on Computer
  Vision},
pp. \bfpage{1960}--\blpage{1970}
(\byear{2021})
\end{bchapter}
\endbibitem

\bibitem[\protect\citeauthoryear{Coffey and
  Vaandering}{2010}]{coffey2010patient}
\begin{barticle}
\bauthor{\bsnm{Coffey}, \binits{M.}},
\bauthor{\bsnm{Vaandering}, \binits{A.}}:
\batitle{Patient setup for pet/ct acquisition in radiotherapy planning}.
\bjtitle{Radiotherapy and Oncology}
\bvolume{96}(\bissue{3}),
\bfpage{298}--\blpage{301}
(\byear{2010})
\end{barticle}
\endbibitem

\bibitem[\protect\citeauthoryear{Mettler~Jr and
  Guiberteau}{2012}]{mettler2012essentials}
\begin{bbook}
\bauthor{\bsnm{Mettler~Jr}, \binits{F.A.}},
\bauthor{\bsnm{Guiberteau}, \binits{M.J.}}:
\bbtitle{Essentials of Nuclear Medicine Imaging: Expert Consult-online and
  Print}.
\bpublisher{Elsevier Health Sciences}, \blocation{???}
(\byear{2012})
\end{bbook}
\endbibitem

\bibitem[\protect\citeauthoryear{Ying et~al.}{2019}]{ying2019x2ct}
\begin{bchapter}
\bauthor{\bsnm{Ying}, \binits{X.}},
\bauthor{\bsnm{Guo}, \binits{H.}},
\bauthor{\bsnm{Ma}, \binits{K.}},
\bauthor{\bsnm{Wu}, \binits{J.}},
\bauthor{\bsnm{Weng}, \binits{Z.}},
\bauthor{\bsnm{Zheng}, \binits{Y.}}:
\bctitle{X2ct-gan: reconstructing ct from biplanar x-rays with generative
  adversarial networks}.
In: \bbtitle{Proceedings of the IEEE/CVF Conference on Computer Vision and
  Pattern Recognition},
pp. \bfpage{10619}--\blpage{10628}
(\byear{2019})
\end{bchapter}
\endbibitem

\bibitem[\protect\citeauthoryear{Ratul et~al.}{2021}]{ratul2021ccx}
\begin{bchapter}
\bauthor{\bsnm{Ratul}, \binits{M.A.R.}},
\bauthor{\bsnm{Yuan}, \binits{K.}},
\bauthor{\bsnm{Lee}, \binits{W.}}:
\bctitle{Ccx-raynet: a class conditioned convolutional neural network for
  biplanar x-rays to ct volume}.
In: \bbtitle{2021 IEEE 18th International Symposium on Biomedical Imaging
  (ISBI)},
pp. \bfpage{1655}--\blpage{1659}
(\byear{2021}).
\bcomment{IEEE}
\end{bchapter}
\endbibitem

\bibitem[\protect\citeauthoryear{Jiang}{2022}]{jiang2022mfct}
\begin{botherref}
\oauthor{\bsnm{Jiang}, \binits{Y.}}:
Mfct-gan: multi-information network to reconstruct ct volumes for security
  screening.
Journal of Intelligent Manufacturing and Special Equipment
(2022)
\end{botherref}
\endbibitem

\bibitem[\protect\citeauthoryear{Sun et~al.}{2022}]{sun2022ultra}
\begin{barticle}
\bauthor{\bsnm{Sun}, \binits{X.}},
\bauthor{\bsnm{Li}, \binits{X.}},
\bauthor{\bsnm{Chen}, \binits{P.}}:
\batitle{An ultra-sparse view ct imaging method based on x-ray2ctnet}.
\bjtitle{IEEE Transactions on Computational Imaging}
\bvolume{8},
\bfpage{733}--\blpage{742}
(\byear{2022})
\end{barticle}
\endbibitem

\bibitem[\protect\citeauthoryear{Shen et~al.}{2022}]{shen2022geometry}
\begin{barticle}
\bauthor{\bsnm{Shen}, \binits{L.}},
\bauthor{\bsnm{Zhao}, \binits{W.}},
\bauthor{\bsnm{Capaldi}, \binits{D.}},
\bauthor{\bsnm{Pauly}, \binits{J.}},
\bauthor{\bsnm{Xing}, \binits{L.}}:
\batitle{A geometry-informed deep learning framework for ultra-sparse 3d
  tomographic image reconstruction}.
\bjtitle{Computers in Biology and Medicine}
\bvolume{148},
\bfpage{105710}
(\byear{2022})
\end{barticle}
\endbibitem

\bibitem[\protect\citeauthoryear{Jin et~al.}{2017}]{jin2017deep}
\begin{barticle}
\bauthor{\bsnm{Jin}, \binits{K.H.}},
\bauthor{\bsnm{McCann}, \binits{M.T.}},
\bauthor{\bsnm{Froustey}, \binits{E.}},
\bauthor{\bsnm{Unser}, \binits{M.}}:
\batitle{Deep convolutional neural network for inverse problems in imaging}.
\bjtitle{IEEE Transactions on Image Processing}
\bvolume{26}(\bissue{9}),
\bfpage{4509}--\blpage{4522}
(\byear{2017})
\end{barticle}
\endbibitem

\bibitem[\protect\citeauthoryear{Shen et~al.}{2022}]{shen2022nerp}
\begin{botherref}
\oauthor{\bsnm{Shen}, \binits{L.}},
\oauthor{\bsnm{Pauly}, \binits{J.}},
\oauthor{\bsnm{Xing}, \binits{L.}}:
Nerp: implicit neural representation learning with prior embedding for sparsely
  sampled image reconstruction.
IEEE Transactions on Neural Networks and Learning Systems
(2022)
\end{botherref}
\endbibitem

\bibitem[\protect\citeauthoryear{Yu et~al.}{2019}]{yu2019ea}
\begin{barticle}
\bauthor{\bsnm{Yu}, \binits{B.}},
\bauthor{\bsnm{Zhou}, \binits{L.}},
\bauthor{\bsnm{Wang}, \binits{L.}},
\bauthor{\bsnm{Shi}, \binits{Y.}},
\bauthor{\bsnm{Fripp}, \binits{J.}},
\bauthor{\bsnm{Bourgeat}, \binits{P.}}:
\batitle{Ea-gans: edge-aware generative adversarial networks for cross-modality
  mr image synthesis}.
\bjtitle{IEEE transactions on medical imaging}
\bvolume{38}(\bissue{7}),
\bfpage{1750}--\blpage{1762}
(\byear{2019})
\end{barticle}
\endbibitem

\bibitem[\protect\citeauthoryear{Ch{\^e}nes and
  Schmid}{2021}]{chenes2021revisiting}
\begin{bchapter}
\bauthor{\bsnm{Ch{\^e}nes}, \binits{C.}},
\bauthor{\bsnm{Schmid}, \binits{J.}}:
\bctitle{Revisiting contour-driven and knowledge-based deformable models:
  Application to 2d-3d proximal femur reconstruction from x-ray images}.
In: \bbtitle{International Conference on Medical Image Computing and
  Computer-Assisted Intervention},
pp. \bfpage{451}--\blpage{460}
(\byear{2021}).
\bcomment{Springer}
\end{bchapter}
\endbibitem

\bibitem[\protect\citeauthoryear{Bera and Biswas}{2021}]{bera2021noise}
\begin{barticle}
\bauthor{\bsnm{Bera}, \binits{S.}},
\bauthor{\bsnm{Biswas}, \binits{P.K.}}:
\batitle{Noise conscious training of non local neural network powered by self
  attentive spectral normalized markovian patch gan for low dose ct denoising}.
\bjtitle{IEEE Transactions on Medical Imaging}
\bvolume{40}(\bissue{12}),
\bfpage{3663}--\blpage{3673}
(\byear{2021})
\end{barticle}
\endbibitem

\bibitem[\protect\citeauthoryear{Wu et~al.}{2022}]{wu2022vessel}
\begin{barticle}
\bauthor{\bsnm{Wu}, \binits{C.}},
\bauthor{\bsnm{Zhang}, \binits{H.}},
\bauthor{\bsnm{Chen}, \binits{J.}},
\bauthor{\bsnm{Gao}, \binits{Z.}},
\bauthor{\bsnm{Zhang}, \binits{P.}},
\bauthor{\bsnm{Muhammad}, \binits{K.}},
\bauthor{\bsnm{Del~Ser}, \binits{J.}}:
\batitle{Vessel-gan: angiographic reconstructions from myocardial ct perfusion
  with explainable generative adversarial networks}.
\bjtitle{Future Generation Computer Systems}
\bvolume{130},
\bfpage{128}--\blpage{139}
(\byear{2022})
\end{barticle}
\endbibitem

\bibitem[\protect\citeauthoryear{Liu et~al.}{2021}]{liu2021ct}
\begin{barticle}
\bauthor{\bsnm{Liu}, \binits{Y.}},
\bauthor{\bsnm{Chen}, \binits{A.}},
\bauthor{\bsnm{Shi}, \binits{H.}},
\bauthor{\bsnm{Huang}, \binits{S.}},
\bauthor{\bsnm{Zheng}, \binits{W.}},
\bauthor{\bsnm{Liu}, \binits{Z.}},
\bauthor{\bsnm{Zhang}, \binits{Q.}},
\bauthor{\bsnm{Yang}, \binits{X.}}:
\batitle{Ct synthesis from mri using multi-cycle gan for head-and-neck
  radiation therapy}.
\bjtitle{Computerized Medical Imaging and Graphics}
\bvolume{91},
\bfpage{101953}
(\byear{2021})
\end{barticle}
\endbibitem

\bibitem[\protect\citeauthoryear{Oulbacha and Kadoury}{2020}]{oulbacha2020mri}
\begin{bchapter}
\bauthor{\bsnm{Oulbacha}, \binits{R.}},
\bauthor{\bsnm{Kadoury}, \binits{S.}}:
\bctitle{Mri to ct synthesis of the lumbar spine from a pseudo-3d cycle gan}.
In: \bbtitle{2020 IEEE 17th International Symposium on Biomedical Imaging
  (ISBI)},
pp. \bfpage{1784}--\blpage{1787}
(\byear{2020}).
\bcomment{IEEE}
\end{bchapter}
\endbibitem

\bibitem[\protect\citeauthoryear{Henzler et~al.}{2018}]{henzler2018single}
\begin{bchapter}
\bauthor{\bsnm{Henzler}, \binits{P.}},
\bauthor{\bsnm{Rasche}, \binits{V.}},
\bauthor{\bsnm{Ropinski}, \binits{T.}},
\bauthor{\bsnm{Ritschel}, \binits{T.}}:
\bctitle{Single-image tomography: 3d volumes from 2d cranial x-rays}.
In: \bbtitle{Computer Graphics Forum},
vol. \bseriesno{37},
pp. \bfpage{377}--\blpage{388}
(\byear{2018}).
\bcomment{Wiley Online Library}
\end{bchapter}
\endbibitem

\bibitem[\protect\citeauthoryear{Jecklin et~al.}{2022}]{jecklin2022x23d}
\begin{barticle}
\bauthor{\bsnm{Jecklin}, \binits{S.}},
\bauthor{\bsnm{Jancik}, \binits{C.}},
\bauthor{\bsnm{Farshad}, \binits{M.}},
\bauthor{\bsnm{F{\"u}rnstahl}, \binits{P.}},
\bauthor{\bsnm{Esfandiari}, \binits{H.}}:
\batitle{X23d—intraoperative 3d lumbar spine shape reconstruction based on
  sparse multi-view x-ray data}.
\bjtitle{Journal of Imaging}
\bvolume{8}(\bissue{10}),
\bfpage{271}
(\byear{2022})
\end{barticle}
\endbibitem

\bibitem[\protect\citeauthoryear{Shen et~al.}{2019}]{shen2019patient}
\begin{barticle}
\bauthor{\bsnm{Shen}, \binits{L.}},
\bauthor{\bsnm{Zhao}, \binits{W.}},
\bauthor{\bsnm{Xing}, \binits{L.}}:
\batitle{Patient-specific reconstruction of volumetric computed tomography
  images from a single projection view via deep learning}.
\bjtitle{Nature biomedical engineering}
\bvolume{3}(\bissue{11}),
\bfpage{880}--\blpage{888}
(\byear{2019})
\end{barticle}
\endbibitem

\bibitem[\protect\citeauthoryear{Gulrajani
  et~al.}{2017}]{gulrajani2017improved}
\begin{botherref}
\oauthor{\bsnm{Gulrajani}, \binits{I.}},
\oauthor{\bsnm{Ahmed}, \binits{F.}},
\oauthor{\bsnm{Arjovsky}, \binits{M.}},
\oauthor{\bsnm{Dumoulin}, \binits{V.}},
\oauthor{\bsnm{Courville}, \binits{A.C.}}:
Improved training of wasserstein gans.
Advances in neural information processing systems
\textbf{30}
(2017)
\end{botherref}
\endbibitem

\bibitem[\protect\citeauthoryear{Liu et~al.}{2020}]{liu2020towards}
\begin{bchapter}
\bauthor{\bsnm{Liu}, \binits{B.}},
\bauthor{\bsnm{Zhu}, \binits{Y.}},
\bauthor{\bsnm{Song}, \binits{K.}},
\bauthor{\bsnm{Elgammal}, \binits{A.}}:
\bctitle{Towards faster and stabilized gan training for high-fidelity few-shot
  image synthesis}.
In: \bbtitle{International Conference on Learning Representations}
(\byear{2020})
\end{bchapter}
\endbibitem

\bibitem[\protect\citeauthoryear{Fan et~al.}{2022}]{fan2022tr}
\begin{botherref}
\oauthor{\bsnm{Fan}, \binits{C.-C.}},
\oauthor{\bsnm{Peng}, \binits{L.}},
\oauthor{\bsnm{Wang}, \binits{T.}},
\oauthor{\bsnm{Yang}, \binits{H.}},
\oauthor{\bsnm{Zhou}, \binits{X.-H.}},
\oauthor{\bsnm{Ni}, \binits{Z.-L.}},
\oauthor{\bsnm{Chen}, \binits{S.}},
\oauthor{\bsnm{Zhou}, \binits{Y.-J.}},
\oauthor{\bsnm{Hou}, \binits{Z.-G.}}, et al.:
Tr-gan: Multi-session future mri prediction with temporal recurrent generative
  adversarial network.
IEEE Transactions on Medical Imaging
(2022)
\end{botherref}
\endbibitem

\bibitem[\protect\citeauthoryear{Jiang et~al.}{2021}]{jiang2021synthesis}
\begin{barticle}
\bauthor{\bsnm{Jiang}, \binits{G.}},
\bauthor{\bsnm{Wei}, \binits{J.}},
\bauthor{\bsnm{Xu}, \binits{Y.}},
\bauthor{\bsnm{He}, \binits{Z.}},
\bauthor{\bsnm{Zeng}, \binits{H.}},
\bauthor{\bsnm{Wu}, \binits{J.}},
\bauthor{\bsnm{Qin}, \binits{G.}},
\bauthor{\bsnm{Chen}, \binits{W.}},
\bauthor{\bsnm{Lu}, \binits{Y.}}:
\batitle{Synthesis of mammogram from digital breast tomosynthesis using deep
  convolutional neural network with gradient guided cgans}.
\bjtitle{IEEE Transactions on Medical Imaging}
\bvolume{40}(\bissue{8}),
\bfpage{2080}--\blpage{2091}
(\byear{2021})
\end{barticle}
\endbibitem

\bibitem[\protect\citeauthoryear{Zhu et~al.}{2017}]{zhu2017unpaired}
\begin{bchapter}
\bauthor{\bsnm{Zhu}, \binits{J.-Y.}},
\bauthor{\bsnm{Park}, \binits{T.}},
\bauthor{\bsnm{Isola}, \binits{P.}},
\bauthor{\bsnm{Efros}, \binits{A.A.}}:
\bctitle{Unpaired image-to-image translation using cycle-consistent adversarial
  networks}.
In: \bbtitle{Proceedings of the IEEE International Conference on Computer
  Vision},
pp. \bfpage{2223}--\blpage{2232}
(\byear{2017})
\end{bchapter}
\endbibitem

\bibitem[\protect\citeauthoryear{Gonz{\'{a}}lez and Woods}{2008}]{GonzalezW08}
\begin{bbook}
\bauthor{\bsnm{Gonz{\'{a}}lez}, \binits{R.C.}},
\bauthor{\bsnm{Woods}, \binits{R.E.}}:
\bbtitle{Digital Image Processing, 3rd Edition}.
\bpublisher{Pearson Education}, \blocation{???}
(\byear{2008})
\end{bbook}
\endbibitem

\bibitem[\protect\citeauthoryear{Lu et~al.}{2010}]{lu2010selective}
\begin{barticle}
\bauthor{\bsnm{Lu}, \binits{Y.}},
\bauthor{\bsnm{Chan}, \binits{H.-P.}},
\bauthor{\bsnm{Wei}, \binits{J.}},
\bauthor{\bsnm{Hadjiiski}, \binits{L.M.}}:
\batitle{Selective-diffusion regularization for enhancement of
  microcalcifications in digital breast tomosynthesis reconstruction}.
\bjtitle{Medical physics}
\bvolume{37}(\bissue{11}),
\bfpage{6003}--\blpage{6014}
(\byear{2010})
\end{barticle}
\endbibitem

\bibitem[\protect\citeauthoryear{Wei et~al.}{2019}]{wei2019synthesizing}
\begin{barticle}
\bauthor{\bsnm{Wei}, \binits{J.}},
\bauthor{\bsnm{Chan}, \binits{H.-P.}},
\bauthor{\bsnm{Helvie}, \binits{M.A.}},
\bauthor{\bsnm{Roubidoux}, \binits{M.A.}},
\bauthor{\bsnm{Neal}, \binits{C.H.}},
\bauthor{\bsnm{Lu}, \binits{Y.}},
\bauthor{\bsnm{Hadjiiski}, \binits{L.M.}},
\bauthor{\bsnm{Zhou}, \binits{C.}}:
\batitle{Synthesizing mammogram from digital breast tomosynthesis}.
\bjtitle{Physics in Medicine \& Biology}
\bvolume{64}(\bissue{4}),
\bfpage{045011}
(\byear{2019})
\end{barticle}
\endbibitem

\bibitem[\protect\citeauthoryear{Stimpel et~al.}{2019}]{stimpel2019projection}
\begin{barticle}
\bauthor{\bsnm{Stimpel}, \binits{B.}},
\bauthor{\bsnm{Syben}, \binits{C.}},
\bauthor{\bsnm{W{\"u}rfl}, \binits{T.}},
\bauthor{\bsnm{Breininger}, \binits{K.}},
\bauthor{\bsnm{Hoelter}, \binits{P.}},
\bauthor{\bsnm{D{\"o}rfler}, \binits{A.}},
\bauthor{\bsnm{Maier}, \binits{A.}}:
\batitle{projection-to-projection translation for hybrid x-ray and magnetic
  resonance imaging}.
\bjtitle{Scientific Reports}
\bvolume{9}(\bissue{1}),
\bfpage{18814}
(\byear{2019})
\end{barticle}
\endbibitem

\bibitem[\protect\citeauthoryear{Kim et~al.}{2016}]{kim2016deeply}
\begin{bchapter}
\bauthor{\bsnm{Kim}, \binits{J.}},
\bauthor{\bsnm{Lee}, \binits{J.K.}},
\bauthor{\bsnm{Lee}, \binits{K.M.}}:
\bctitle{Deeply-recursive convolutional network for image super-resolution}.
In: \bbtitle{Proceedings of the IEEE Conference on Computer Vision and Pattern
  Recognition},
pp. \bfpage{1637}--\blpage{1645}
(\byear{2016})
\end{bchapter}
\endbibitem

\bibitem[\protect\citeauthoryear{Zhang et~al.}{2018}]{cvpr/ZhangIESW18}
\begin{bchapter}
\bauthor{\bsnm{Zhang}, \binits{R.}},
\bauthor{\bsnm{Isola}, \binits{P.}},
\bauthor{\bsnm{Efros}, \binits{A.A.}},
\bauthor{\bsnm{Shechtman}, \binits{E.}},
\bauthor{\bsnm{Wang}, \binits{O.}}:
\bctitle{The unreasonable effectiveness of deep features as a perceptual
  metric}.
In: \bbtitle{2018 {IEEE} Conference on Computer Vision and Pattern Recognition,
  {CVPR} 2018},
pp. \bfpage{586}--\blpage{595}.
\bpublisher{Computer Vision Foundation / {IEEE} Computer Society},
  \blocation{???}
(\byear{2018})
\end{bchapter}
\endbibitem

\bibitem[\protect\citeauthoryear{Tran et~al.}{2021}]{tran2021data}
\begin{barticle}
\bauthor{\bsnm{Tran}, \binits{N.-T.}},
\bauthor{\bsnm{Tran}, \binits{V.-H.}},
\bauthor{\bsnm{Nguyen}, \binits{N.-B.}},
\bauthor{\bsnm{Nguyen}, \binits{T.-K.}},
\bauthor{\bsnm{Cheung}, \binits{N.-M.}}:
\batitle{On data augmentation for gan training}.
\bjtitle{IEEE Transactions on Image Processing}
\bvolume{30},
\bfpage{1882}--\blpage{1897}
(\byear{2021})
\end{barticle}
\endbibitem

\bibitem[\protect\citeauthoryear{Liang et~al.}{2022}]{liang2022sketch}
\begin{barticle}
\bauthor{\bsnm{Liang}, \binits{J.}},
\bauthor{\bsnm{Yang}, \binits{X.}},
\bauthor{\bsnm{Huang}, \binits{Y.}},
\bauthor{\bsnm{Li}, \binits{H.}},
\bauthor{\bsnm{He}, \binits{S.}},
\bauthor{\bsnm{Hu}, \binits{X.}},
\bauthor{\bsnm{Chen}, \binits{Z.}},
\bauthor{\bsnm{Xue}, \binits{W.}},
\bauthor{\bsnm{Cheng}, \binits{J.}},
\bauthor{\bsnm{Ni}, \binits{D.}}:
\batitle{Sketch guided and progressive growing gan for realistic and editable
  ultrasound image synthesis}.
\bjtitle{Medical Image Analysis}
\bvolume{79},
\bfpage{102461}
(\byear{2022})
\end{barticle}
\endbibitem

\bibitem[\protect\citeauthoryear{Sara et~al.}{2019}]{sara2019image}
\begin{barticle}
\bauthor{\bsnm{Sara}, \binits{U.}},
\bauthor{\bsnm{Akter}, \binits{M.}},
\bauthor{\bsnm{Uddin}, \binits{M.S.}}:
\batitle{Image quality assessment through fsim, ssim, mse and psnr—a
  comparative study}.
\bjtitle{Journal of Computer and Communications}
\bvolume{7}(\bissue{3}),
\bfpage{8}--\blpage{18}
(\byear{2019})
\end{barticle}
\endbibitem

\bibitem[\protect\citeauthoryear{Shcherbakov
  et~al.}{2013}]{shcherbakov2013survey}
\begin{barticle}
\bauthor{\bsnm{Shcherbakov}, \binits{M.V.}},
\bauthor{\bsnm{Brebels}, \binits{A.}},
\bauthor{\bsnm{Shcherbakova}, \binits{N.L.}},
\bauthor{\bsnm{Tyukov}, \binits{A.P.}},
\bauthor{\bsnm{Janovsky}, \binits{T.A.}},
\bauthor{\bsnm{Kamaev}, \binits{V.A.}}, \betal:
\batitle{A survey of forecast error measures}.
\bjtitle{World applied sciences journal}
\bvolume{24}(\bissue{24}),
\bfpage{171}--\blpage{176}
(\byear{2013})
\end{barticle}
\endbibitem

\bibitem[\protect\citeauthoryear{Caponetti and Fanelli}{1990}]{caponetti19903d}
\begin{bchapter}
\bauthor{\bsnm{Caponetti}, \binits{L.}},
\bauthor{\bsnm{Fanelli}, \binits{A.}}:
\bctitle{3d bone reconstruction from two x-ray views}.
In: \bbtitle{[1990] Proceedings of the Twelfth Annual International Conference
  of the IEEE Engineering in Medicine and Biology Society},
pp. \bfpage{208}--\blpage{210}
(\byear{1990}).
\bcomment{IEEE}
\end{bchapter}
\endbibitem

\bibitem[\protect\citeauthoryear{Choy et~al.}{2016}]{choy20163d}
\begin{bchapter}
\bauthor{\bsnm{Choy}, \binits{C.B.}},
\bauthor{\bsnm{Xu}, \binits{D.}},
\bauthor{\bsnm{Gwak}, \binits{J.}},
\bauthor{\bsnm{Chen}, \binits{K.}},
\bauthor{\bsnm{Savarese}, \binits{S.}}:
\bctitle{3d-r2n2: A unified approach for single and multi-view 3d object
  reconstruction}.
In: \bbtitle{European Conference on Computer Vision},
pp. \bfpage{628}--\blpage{644}
(\byear{2016}).
\bcomment{Springer}
\end{bchapter}
\endbibitem

\bibitem[\protect\citeauthoryear{Mirza et~al.}{2014}]{mirza2014generative}
\begin{barticle}
\bauthor{\bsnm{Mirza}, \binits{M.}},
\bauthor{\bsnm{Xu}, \binits{B.}},
\bauthor{\bsnm{Warde-Farley}, \binits{D.}},
\bauthor{\bsnm{Ozair}, \binits{S.}},
\bauthor{\bsnm{Courville}, \binits{A.}},
\bauthor{\bsnm{Bengio}, \binits{Y.}},
\bauthor{\bsnm{Goodfellow}, \binits{I.J.}},
\bauthor{\bsnm{Pouget-Abadie}, \binits{J.}}:
\batitle{Generative adversarial nets}.
\bjtitle{Proceedings of the Advances in Neural Information Processing Systems}
\bvolume{27},
\bfpage{2672}--\blpage{2680}
(\byear{2014})
\end{barticle}
\endbibitem

\bibitem[\protect\citeauthoryear{Kingma~Diederik and
  Adam}{2014}]{kingma2014method}
\begin{botherref}
\oauthor{\bsnm{Kingma~Diederik}, \binits{P.}},
\oauthor{\bsnm{Adam}, \binits{J.B.}}:
A method for stochastic optimization.
arXiv preprint arXiv:1412.6980
(2014)
\end{botherref}
\endbibitem

\end{thebibliography}

\section*{Acknowledgement}{
This research is supported by the Guangdong Provincial Key-Area Research and Development Program  (2022B0101010005), Qinghai Provincial Science and Technology Research Program (2021-QY-206), National Natural Science Foundation of China (62071201), and Guangdong Basic and Applied Basic Research Foundation (No. 2022A1515010119). 
}

\end{document}